\documentclass[twocolumn]{aastex631}

\newcommand{\FRB}{{\rm FRB}\,20180916{\rm B}}
\newcommand{\phase}{\overline{\phi}}

\usepackage{graphicx} 
\usepackage{rotating}
\usepackage{longtable}

\shortauthors{Bhattacharyya et al.}
\graphicspath{{./}{figures/}}
\usepackage{gensymb}
\begin{document}

\title{Wideband Monitoring of FRB 20180916B Across a Half-Decade Bandwidth Using the Upgraded GMRT}
\author[0000-0003-0669-873X]{Siddhartha Bhattacharyya}
\affiliation{National Centre for Radio Astrophysics,
Tata Institute of Fundamental Research,
Pune, 411007, India}
\author[0000-0002-2892-8025]{Jayanta Roy}
\affiliation{National Centre for Radio Astrophysics,
Tata Institute of Fundamental Research,
Pune, 411007, India}
\author[0000-0002-2864-4110]{Apurba Bera}
\affiliation{International Centre for Radio Astronomy Research (ICRAR),
Curtin University,
Bentley, WA 6102, Australia}
\correspondingauthor{Siddhartha Bhattacharyya}
\email{sbhattacharyya@ncra.tifr.res.in, sid.phy.in@gmail.com}
\begin{abstract}
With the uGMRT having unprecedented sensitivity and unique capability of providing instantaneous frequency coverage of 250$-$1460 MHz, we studied ${\rm FRB}\,20180916$B over four months sampling during its active phase.
We report the detection of $74$ bursts at Band-3 (i.e. 250$-$500 MHz) and $4$ bursts at Band-4 (i.e. 550$-$750 MHz) of uGMRT providing a burst rate of $\sim 4$ bursts/hour above a fluence of 0.05 Jy ms.
We find that the source emits maximum energy and luminosity up to a fractional bandwidth of 70 MHz near the middle of its activity window consistent with earlier studies. 
We see a strong correlation between the excess dispersion measure and excess scattering width. 
We find that the normalized cumulative distribution of the waiting time can be well-fitted by an exponential function, indicating a stochastic emission process.
We also notice that the cumulative burst rate changes rapidly with the intrinsic energy of the bursts near the middle of the activity window considering the full observed window of 0.4-0.6 of this FRB, where this change is much steeper for the high-energy bursts and shallower for the low-energy bursts.
\end{abstract}
\keywords{methods: observational – techniques: transients; fast radio bursts.}
\section{Introduction} \label{sec:1}
Fast radio bursts (FRBs) are short-duration, bright, dispersed radio pulses of yet unknown physical origin \citep{rajwade2024}. 
For FRBs, the dispersion measure ($DM$) values are typically $5$ to $20$ times larger than the Milky Way's contribution to the observed DMs ($DM_{MW}$), implying that FRBs are extragalactic in origin and located at cosmological distances, corresponding to a redshift range of 0.09–3.5 \citep{rajwade2024}. 
For more details about the physical properties of FRBs, the reader may refer to the review articles (eg. \citealt{petroff19}).
Several emission models (e.g. \citealt{zhang2023}) involving progenitors from cosmic strings to compact objects have been proposed to explain all the bursts originating from the sources but none of them gives us a clear picture to date. 
One of the main challenges behind the difficulties in constraining the origin of the FRBs is that, in most cases,  they are one-off events, exhibiting a single millisecond burst of radio emission. 

More than $800$ FRBs have been reported to date\footnote{\url{https://www.wis-tns.org/}}.
Approximately 55 FRBs have been found to repeat, with multiple bursts recorded from a single source, leading them to be labelled as ``repeaters.''
Till now only $2$ repeaters, \textit{viz.} ${\rm FRB}\,20121102A$ and $\FRB$ have shown  periodicity (e.g. \citealt{rajwade2020,cruces2021,chime2020a}) in their activities  \citep{braga2024,andersen2019}. 
Further, out of $55$ repeaters detected to date, only $18$ events were localised with their associated host galaxies (refer to https://www.frb-hosts.org/).  
On the other hand, the majority of detected FRBs ($>800$) are \say{one-off} events, where only a single burst has been observed from each source.
However, it is not clear whether the progenitors of repeaters have different physical origins compared to those of one-off events. 
Moreover, the detection of $\sim500$ times fainter bursts from FRB 20171019 \citep{kumar19a} compared to the originally detected burst suggests that the identification of the repeating nature of an FRB may strongly depend on instrumental sensitivity. 
Also, the spectra of many repeaters are not well constrained due to the lack of appropriate beam shape information in some cases (e.g. \citealt{connor2020}).
Considering those facts there is a possibility that all FRBs are repeaters \citep{caleb2019}, and for the one-off, the less sensitive follow-up failed to detect the fainter repeating bursts from the same FRB.

Over the past decade, most FRBs have been detected over a frequency range of $110$ MHz to $8$ GHz (e.g. \citealt{pleunis2021}, \citealt{bethapudi23}) with the only few events from ${\rm FRB}\,20121102A$ also being detected in the $4-8$ GHz frequency range \citep{gajjar18}. 
Though FRB emission extends to lower frequencies, only a few FRB searches were conducted below the frequency of 400 MHz (e.g. \citealt{deneva2016}, \citealt{rajwade2020}, \citealt{pastor2021}, \citealt{pilia2020}). 

Observations of FRBs at low frequencies are important to check whether there is any intrinsic cutoff or turnover frequency in their emission,  which could provide valuable insights into their progenitor models (e.g. \citealt{houben2019}). 
It also may happen that FRBs are undetectable at low frequencies due to (a) spectral turnover arising from propagation effect (e.g. \citealt{chawla20}), and/or (b) multi-path propagation of signals in the intervening medium which is a phenomenon called as scattering (e.g. \citealt{ocker2022}). The facts mentioned above are useful to constrain the properties of circumbursts medium of an FRB detected at low-frequency (e.g. \citealt{chime2020a}, \citealt{ravi19}).
Moreover, the frequency-dependent activity of FRBs \citep{pleunis2021} can also be studied using the observations at low frequencies.

The \citealt{chime2021} reported the detection of a relatively bright repeater $\FRB$.
Later \citealt{marcote20} localized this repeater and found its host galaxy using the European VLBI Network (EVN) taking advantage of its repeating nature.
They found that this repeater is located in a nearby massive star-forming spiral galaxy at a distance of $\sim150$ Mpc ($z\approx0.03$), where this galaxy has $\sim 100$ times larger stellar mass and $\sim 5$ times higher metallicity compared to the dwarf host galaxy associated with the ${\rm FRB}\,20121102A$ \citep{marcote20}. 
Further, no persistent radio source with flux density $> 18\,{\rm \mu Jy}$ at $1.6$ GHz \citep{marcote20}, has been seen at the location of this FRB which implies that if any persistent radio emission is associated with this repeater then its luminosity is a few hundred times lower than that of the persistent source coincident with other repeaters, say ${\rm FRB}\,20121102A$ \citep{marcote2017}.
\citealt{andersen2019} reported the detection of $28$ bursts from this repeater over a time span of $\sim 400$ days and they estimated periodicity of $\sim 16.35$ days in the repetition pattern of this FRB with an active window of $\sim 4$ days. 
The average burst rate of this repeater within the active window is $\sim 1.8$ burst per hour above fluence limit of $5$ Jy ms \citep{chime2020a}, however, this rate is lower towards the edge in comparison to the centre of the active window (e.g. \citealt{marcote20}). 
Later \citealt{chawla20} reported detection of this FRB emission up to 110 MHz using  simultaneous observations with a few partial overlapped bands at LOFAR ($110-190$ MHz), GBT ($300-400$ MHz) and CHIME ($400-800$ MHz).
They also placed a lower limit of the spectral index of this repeater is $\alpha>-1$. 
Recently, \citealt{bethapudi23} reported the detection of eight bursts from this $\FRB$ using the $100-$m Effelsberg radio telescope with frequency range $4-8$ GHz. 
They found clear evidence of chromaticity, as the activity window of this repeater shifts with frequency. The bursts detected at $\sim 6$ GHz \citep{bethapudi23} are narrower in time, wider in frequency, and exhibit lower fractional bandwidths compared to bursts at lower frequencies. 
Additionally, the burst rate varies with frequency as $f^{-2.6 \pm 0.2}$, and it also differs from cycle to cycle, with the spectral index estimated from the rate–frequency relation being $-1.04 \pm 0.08$ \citep{bethapudi23}. 

Previously \citealt{marthi20} reported the detection of 15 bursts from this repeater using the upgraded Giant Metrewave Radio Telescope (uGMRT) over a frequency range of $550–750$ MHz. 
They observed these bursts over three successive cycles of approximately 2 hours each, during the peak of its expected active period. 
The burst rate was found to be highly variable, with the faintest burst being only about 10 times brighter than the brightest Galactic burst detected to date. 
They did not detect any persistent radio source associated with this repeater and placed a 3$\sigma$ upper limit on the persistent radio flux density of 66 $\mu$Jy at 650 MHz.
Further, \citet{sand22} and \citet{trudu23} used the uGMRT to observe this repeater over a frequency range of $250-500$ MHz, detecting four and seven bursts, respectively.
In addition to the above mentioned observations, recently \citet{bethapudi24} reported $116$ bursts from this repeater using uGMRT over a frequency range of $550-750$ MHz.

Here we report the detection of low-frequency radio bursts from $\FRB$ with the uGMRT over a frequency range of 250$-$750 MHz.
Section 2 discusses the details of the observation and data analysis. 
In Section 3, we present our findings on the variation of physical parameters of the detected bursts from this repeater. 
Finally, we discuss and summarize our key findings in Section 4. 
\begin{figure*}
\centering
\includegraphics[width=\textwidth]{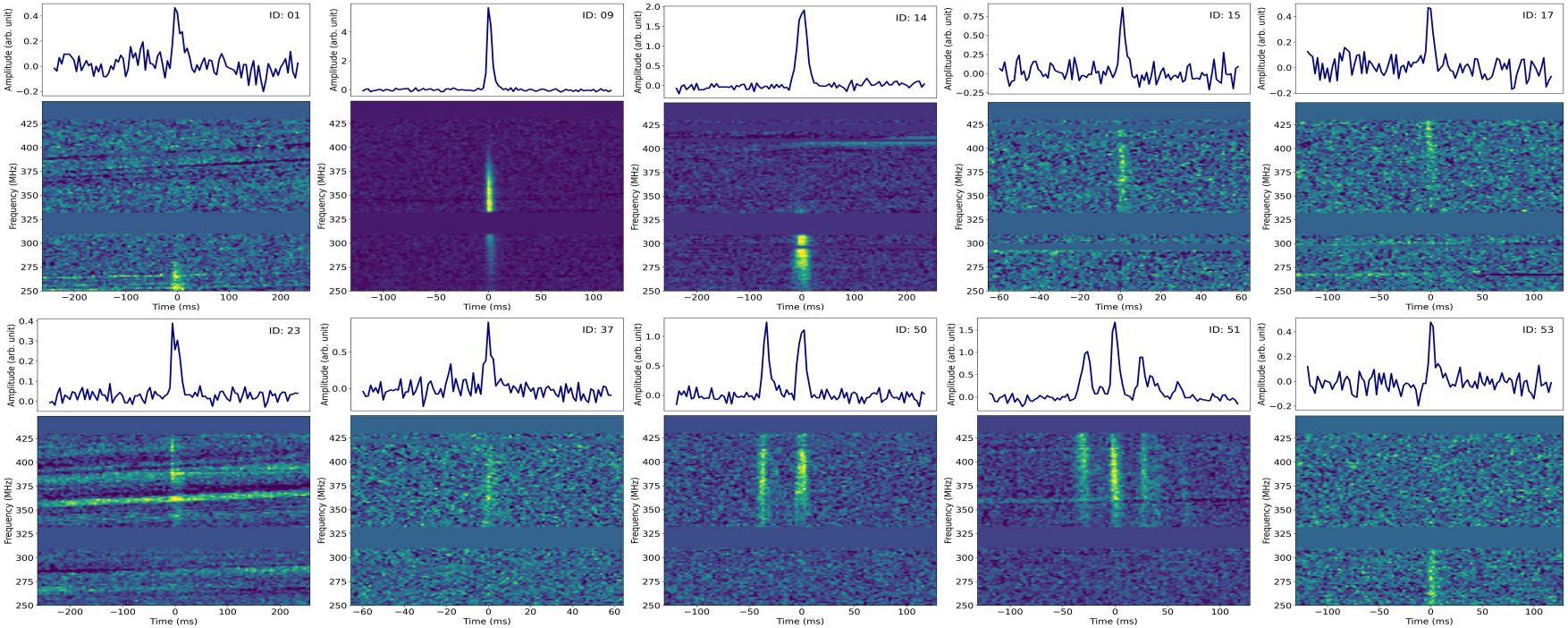}
\caption{A sample gallery of dynamic spectra for 10 bursts from $\FRB$ is shown here and the full set is available at the following link: \url{https://drive.google.com/file/d/1YeVs7i9oJW2MJZvi6Y6KOFnIOX6FkWRk/view?usp=sharing}.}
\label{fig:frb_gallery}
\end{figure*}
\section{Observation and Data Analysis}\label{sec:2}
The energy emitted from an FRB can vary significantly with frequency, highlighting the importance of investigating its emission across a wide frequency range. 
For FRB 20180916B, in addition to the CHIME ($600$ MHz) where the repeater $\FRB$ had been first observed, several other telescopes, viz. LOFAR ($150$ MHz), uGMRT ($650$ MHz), SRT ($328$ MHz), GBT ($350$ MHz), Apertif ($1370$ MHz), Effelsberg ($1420$ MHz), EVN ($1700$ MHz) etc, detected bursts from this repeater. 
Interestingly there is almost no existing observation with a single telescope which probed the bursts over an instantaneous wide frequency range. The current uGMRT observation presented in this paper utilises the wide bandwidth capabilities of uGMRT \citep{gupta17} to observe $\FRB$ simultaneously at $250-500$ MHz (Band-3), $550-850$ MHz (Band-4) and $1060-1460$ MHz (Band-5) respectively. 

\subsection{Observation}\label{sec:2.1}
The uGMRT \citep{gupta17} is an interferometric array of $30$ parabolic dishes each with $45$m in diameter. 
We observed with phased array beams pointed at the position of $\FRB$, i.e. ${\rm RA}=01^{\rm h}58^{\rm m}00^{\rm s}.75$ and ${\rm DEC}=65\degree 43'00''.75$ \citep{marcote20}, in seven epochs over the period from $26$th November $2020$ (${\rm MJD}=59179$) to $21$st March $2021$ (${\rm MJD}=59294$). The multi-subarray observing capability of the uGMRT enabled us to simultaneously form three phased array beams, one at each band $-$ Band-3 (having a usable band of $250-500$ MHz), Band-4 (having a usable band of $550-850$ MHz) and Band-5 (having a usable band of $1060-1460$ MHz) $-$ covering a wide frequency range of $250-1460$ MHz. 
This multi-subarray observing mode splits the full interferometric array of $30$ dishes into three  sub-arrays of $10$ dishes each, which gave us the $10\sigma$ theoretical fluence sensitivity of $0.21$, $0.46$ and $0.38$ Jy ms for Band-3, Band-4 and Band-5 respectively, where to estimate these values we have considered the maximum scattered pulse for both Band-3 and Band-4 and for Band-5 we have assumed the scattering width of 1 ms.  
Each of the phased array beams was recorded in $8192$ frequency channels with $\sim 48.83$ kHz frequency resolution and $655.36$ $\mu$s time resolution.

\subsection{Data Processing}\label{sec:2.2}
The phased array total intensity data with $\sim 48.83$ kHz frequency resolution and $655.36$ $\mu$s time resolution from each sub-array, were processed in the same manner. 
We used the GMRT Pulsar Tool named as \textit{GPTOOL}\footnote{\url{https://github.com/chowdhuryaditya/gptool}} to perform mitigation of radio frequency interference (RFI).
We generated RFI-cleaned \textit{SIGPROC}\citep{sigproc} filterbank file for further processing.
The data were then incoherently dedispersesed  \citep{lorimer2005} to correct for the dispersion caused by the cold ionised plasma in the intervening medium, including the interstellar medium (ISM) of the Milky Way, the host galaxy of the source, and the intergalactic medium (IGM). 
We performed the above-mentioned incoherent dedispersion using the \textit{prepsubband} tool, which is a part of the software \textit{PRESTO}\footnote{\url{https://github.com/scottransom/presto}}, for a $DM$ range of $340-360$ pc cm$^{-3}$ with $DM$ step of $0.01$ pc cm$^{-3}$. 
We then searched for the FRB pulses using \textit{single$\_$pulse$\_$search.py} tool of \textit{PRESTO} above a signal-to-noise ($S/N$) threshold of 5 with an upper limit of pulse width as $65.536$ $ ms$ (i.e. up to box-car of 100 bins).
The lowering of the threshold reduces the chance of missing the band-limited bursts from the repeating FRB. We identified the genuine bursts by manually examining the pulsed emission pattern (potentially band-limited) in the dynamic spectrum and the bow-tie structure in the DM-time plane. We performed this feature extraction of every detection using your$\_$candmaker which is a tool of the software package your\footnote{\url{https://github.com/thepetabyteproject/your}}.
We identified a total of $74$ bursts from $\FRB$ observed at Band-3, only $4$ bursts from at Band-4 and no bursts from Band-5. 
However, we did not find any simultaneous detections in Band-3 and Band-4. 
Out of $74$ bursts detected at Band-3, $26$ bursts were detected with $6.5\leq S/N < 7$, $22$ bursts were detected with $7\leq S/N < 10$ and $26$ bursts were detected with $S/N > 10$ whereas in Band-4 all the four bursts were detected with $S/N > 7$.
For further statistical analysis of burst morphology and FRB parameters, we focus on the bursts detected in Band-3, i.e. a total of $74$ bursts with $S/N\geq 6.5$. 
Figure~\ref{fig:frb_gallery} shows the dynamic spectra along with pulse profiles for all 74 pulses detected from $\FRB$.

\subsection{Activity Phase}\label{sec:2.3}
Previously several studies (e.g. \citealt{chawla20}) reported the activity period of this repeater which is $\sim 16.35$ days with an active window of $\sim 4$ days . 
We have folded the MJDs of the observation epochs and found the phase ($\phi$) of each detected burst from this repeater using the relation 
\begin{equation}
\phi = \frac{T-T_0}{P} - {\rm int}\left(\frac{T-T_0}{P}\right)
\label{eq:phase}
\end{equation}
where $T$ is the MJD of the bursts, $T_0 = 58369.18$ \citep{chawla20} is the starting MJD and $P=16.35$ days \citep{chawla20} is the activity period of $\FRB$. 
\begin{figure}
\centering
\includegraphics[width=\columnwidth]{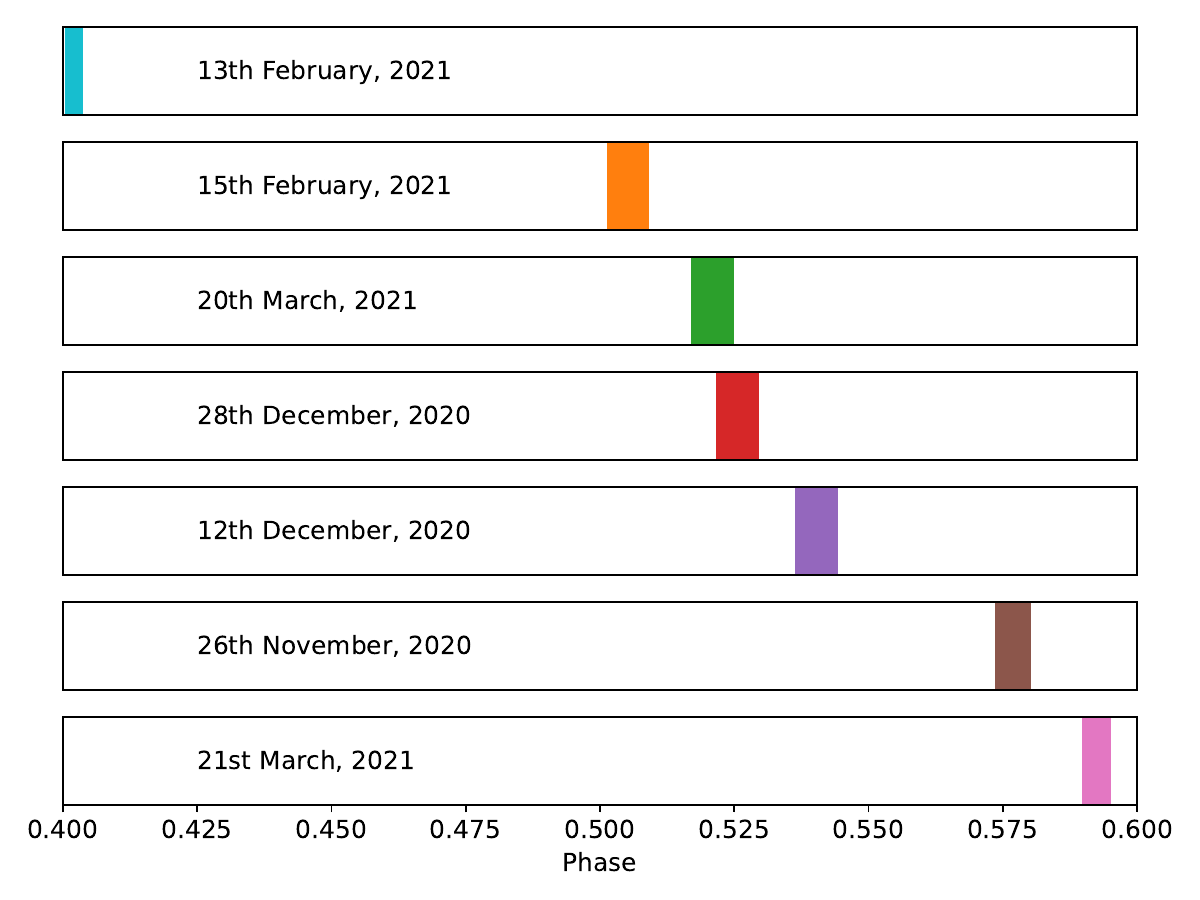}
\caption{The phase space sampling of $\FRB$ was conducted using approximately 18 hours of total uGMRT observations, divided across seven different epochs.}
\label{fig:phase_sample}
\end{figure}
Figure~\ref{fig:phase_sample} shows the phase values (eq.~\ref{eq:phase}) covered in each observation epoch discussed earlier.  
Out of total $\sim 18$ hours of total observation duration (splits into seven different epochs), around $12$ hours observation time (splits into four different epochs) sampled the activity phase of this repeater over a phase range of $\sim 0.51$ to $\sim 0.55$ whereas the full observation duration sampled the phase space from $\phi\simeq 0.4$ to $\simeq 0.6$. 
However, these observations did not uniformly sample the phase space of this repeater within the phase range $0.4 \leq \phi \leq 0.6$, where Figure~\ref{fig:phase_sample} shows a large discontinuity in the phase sampling from $\phi\simeq 0.4$ to $\phi\simeq 0.5$.

The phase values covered by each of the observation epochs are tabulated in Tables~\ref{tab:burst_parameters_1} and \ref{tab:burst_parameters_2}. 
Furthermore, we have not tested the active window of this repeater because the observations referenced above did not cover the repeater's full phase space. 
We have estimated the phase value of all the detected bursts from this repeater using Eq.~\ref{eq:phase} and examined how the physical parameters of the bursts change with their phase values, which is discussed later in this paper.

By combining all seven epochs, we sampled the activity phase of $\FRB$ from approximately $0.4$ to $0.6$ (Table~\ref{tab:literature_P1}), although this phase space sampling was not continuous. It has been reported that the activity phase of this repeater depends on the observing frequency \citep{pastor2021}. 
Therefore, we consider only other observations with frequency ranges comparable to our observing bands. 
These observations include GBT ($300-400$ MHz) \citep{chawla20}, CHIME ($400-800$ MHz) \citep{mckinven23,pearlman20,sand22}, SRT ($300-400$ MHz) \citep{trudu23}, and previous uGMRT ($300-500$ MHz) observations \citep{sand22,trudu23}. 
These studies sampled the activity phase of $\FRB$ from approximately $0.37$ to $0.65$ (Table~\ref{tab:literature_P1}), which is nearly equivalent to our sampling range.

\subsection{Phase Optimised DM}\label{sec:2.4}
The \textit{single$\_$pulse$\_$search} algorithm (discussed earlier) gives us an approximate value of $DM$ of the pulse which further needs a correction depending on the maximization of either the $S/N$ or the power of the detected pulse, where the detection process focuses on enhancing the detectability of the burst by maximising the contrast between the signal and the background noise using $S/N$-optimised $DM$, for studying the intrinsic properties of the burst one needs to preserve the signal's total energy by phase-optimized $DM$. 
First, we discuss the estimation of $DM$ depending on maximising the $S/N$ value. 
During the detection process, we have de-dispersed the detected pulse with different search-$DM$ values and estimated their $S/N$ values. 
We then fit the $S/N$ values of the pulse for different $DM$s with a Gaussian function and estimate the correct $DM$ for which the $S/N$ is maximum which is called the $S/N$-optimised $DM$.

\begin{figure}
\centering
\includegraphics[width=\columnwidth]{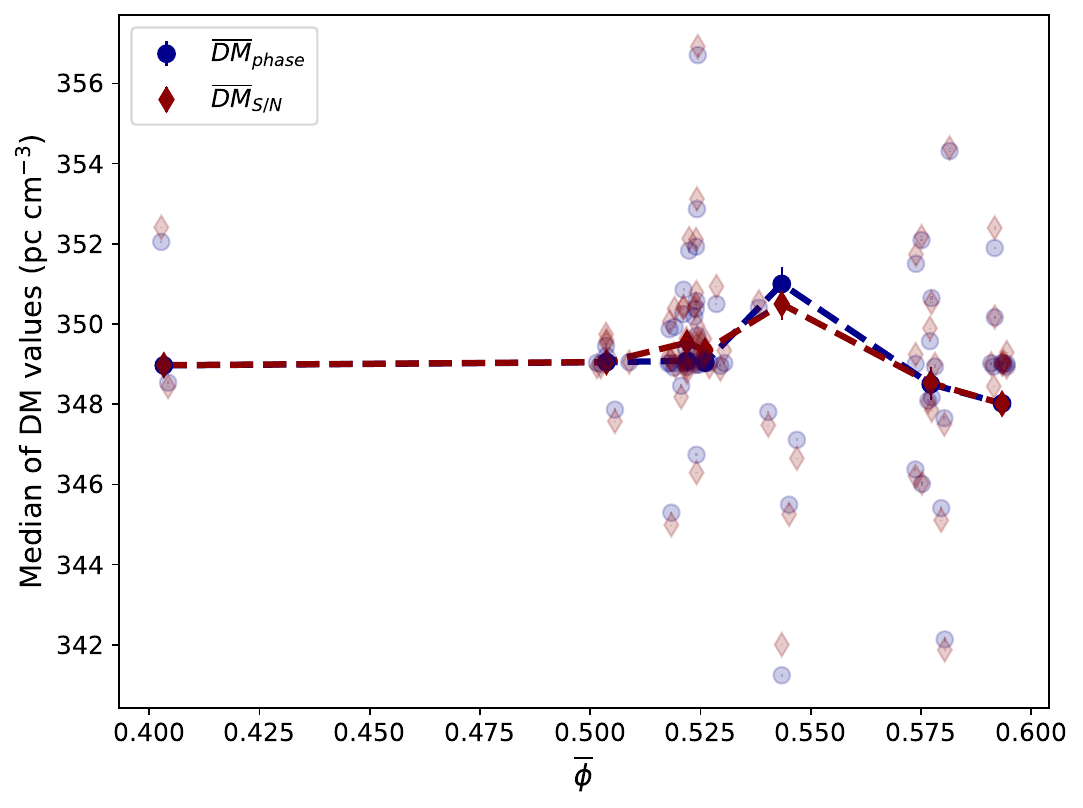}
\caption{The phase optimised and $S/N$ optimised $DM$ values of all 74 bursts, along with the variation of their median values ($\overline{DM}_{phase}$ and $\overline{DM}_{S/N}$), are shown as a function of the median activity phase ($\phase$) of the $\FRB$ at each observation epoch. The error bars represent the uncertainties in the estimates of $\overline{DM}_{phase}$, $\overline{DM}_{S/N}$, and $\phase$.}
\label{fig:phase_DMpow_DMsnr}
\end{figure}

We now discuss the estimation of $DM$ using the $DM_{phase}$\footnote{\url{https://github.com/danielemichilli/DM_phase}} method, which determines the optimal dispersion measure by maximizing the coherent power of a radio signal rather than its intensity. This method \citep{seymour2019} is particularly robust for estimating the best-fit $DM$ in bursts with complex temporal structures and in the presence of interference.

We considered studying the median value per scan of both phase-optimized $DM$, denoted as $\overline{DM}_{phase}$ and $S/N$-optimized $DM$, denoted as $\overline{DM}_{S/N}$ with the median value of activity phases ($\phase$) of this repeater covered at different observation epochs.  
The blue and red points of Figure~\ref{fig:phase_DMpow_DMsnr} show the variation of $\overline{DM}_{phase}$ and $\overline{DM}_{S/N}$ of the detected bursts with $\phase$ of $\FRB$.
We find that for most of the cases the values of $\overline{DM}_{phase}$ and $\overline{DM}_{S/N}$ are almost similar and only for a few cases do they have a very small difference. 
This is quite obvious as for most of the detected bursts presented in this paper we have single component profiles for which phase-optimized $DM$ and $S/N$-optimized $DM$ are almost equal and they are widely differed for the multi-component profile and drifted pulse, which is merely present in the sample of $74$ detected bursts mentioned earlier.  
However, from now and onwards, we use phase-optimized $DM$ as the measured $DM$ values of these $74$ bursts detected at Band-3 of uGMRT. 
The phase-optimized $DM$ values of the detected bursts for each of the observation epochs are tabulated in Tables~\ref{tab:burst_parameters_1} and \ref{tab:burst_parameters_2}.

Combining all seven observing epochs, the $DM$ values of the 74 bursts from this repeater lie within the range of $341.24-356.71$ pc cm$^{-3}$. 
However, the $DM$ values reported from other observations (Table~\ref{tab:literature_P1}) lie within the range of $347.3–370.4$ pc cm$^{-3}$, which overlap with the $DM$ values from our observations, including the detection of a few bursts with lower $DM$ values.

\subsection{Profile and Spectra}\label{sec:2.5}
\begin{figure}
\centering
\includegraphics[width=\columnwidth]{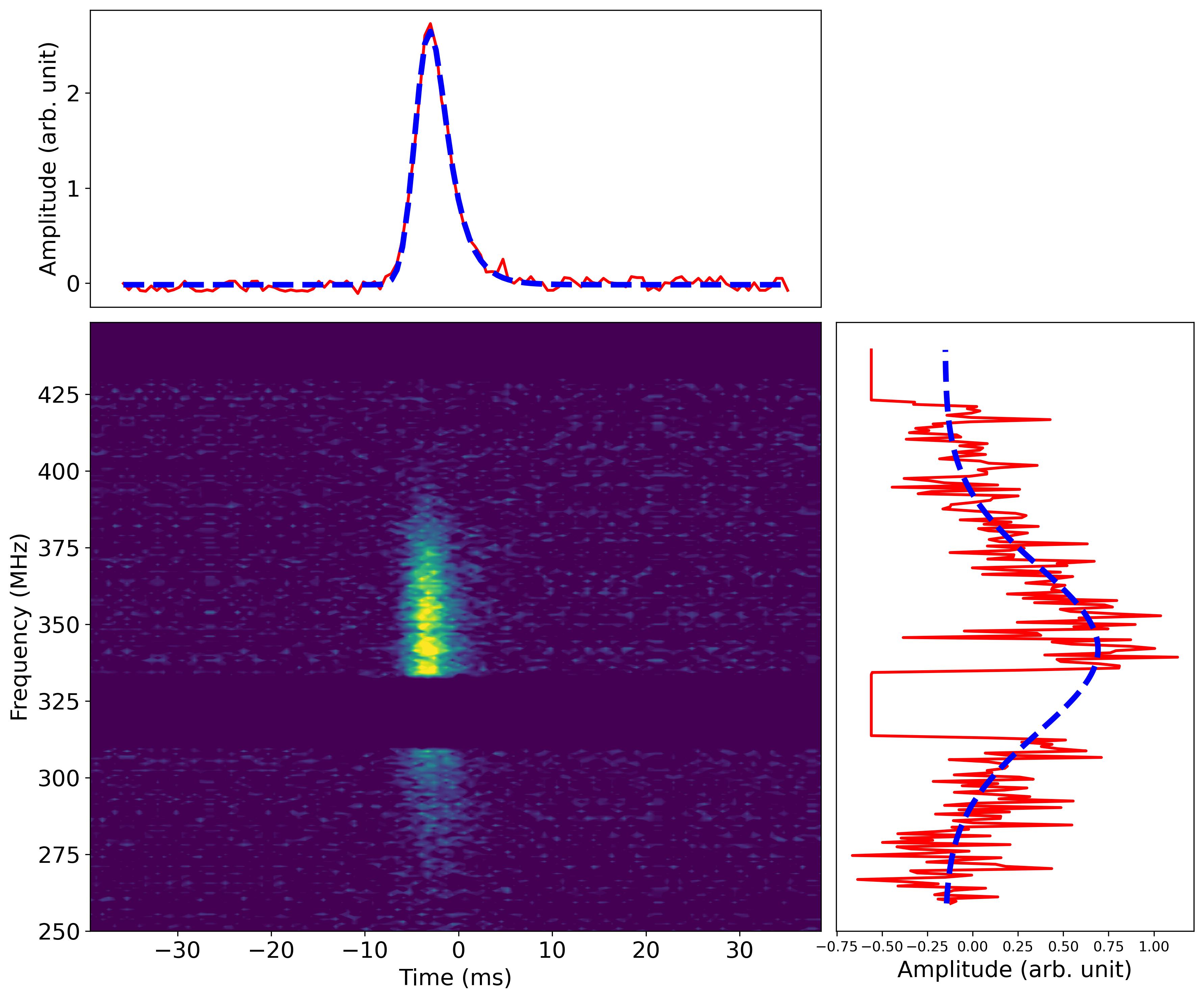}
\caption{The dynamic spectrum (middle panel), profile (upper panel) and spectra (right panel) of a burst from $\FRB$ detected at Band-3 of uGMRT.}
\label{fig:sample_burst}
\end{figure}

We will now discuss the profile and spectra fitting of the detected bursts.
The middle panel of Figure~\ref{fig:sample_burst} shows the colour scale plot of the filterbank file, commonly known as waterfall diagram, of a detected burst, where the X-axis is time of the observation and Y-axis is the frequency of the observation. 
The signal is present in band-3 of the uGMRT starting from 300 MHz. 
The colour gradient of the waterfall diagram (Figure~\ref{fig:sample_burst}) shows the intensity of the burst in the frequency-time plane. 
On the upper and right panels we have the profile of that burst by averaging over the frequency and the spectra of that burst by averaging over the time duration respectively. 
The estimated profile and spectra have been shown in the red lines of the upper and right panels of the Figure~\ref{fig:sample_burst} respectively. 

The pulse profile of the burst is fitted with a Gaussian function convolved with an exponential tail, which is given by
\begin{equation}
I_{P}(t) = A_P \exp\left(-\frac{t-\mu_P}{\tau}\right) {\rm Erfc} \left(-\frac{t-\mu_P-(\sigma_P^2/\tau)}{\sigma_P\sqrt{2}}\right) 
\label{eq:profile_fit}
\end{equation}
where, $I_P(t)$ is intensity of the pulse profile for a given time $t$, and the fitting parameters $A_P$, $\mu_P$, $\sigma_P$ and $\tau$ are determined by the least square method.  
We have estimated the intrinsic and scattering widths of a burst using Eq.~\ref{eq:profile_fit}. 
The parameter $\sigma_P\times 2\ln 2$ gives us the intrinsic width ($w_{int}$) of the pulse for which the intensity of the pulse is dropped by $50\%$ of its peak value and the parameter $\tau$ gives us the scattering width respectively. 
The blue dashed line in the upper panel of Figure~\ref{fig:sample_burst} shows the fitting of the pulse profile using Eq.~\ref{eq:profile_fit}. 
The values of intrinsic and scattering width of the detected bursts for each of the observation epochs are tabulated in Tables~\ref{tab:burst_parameters_1} and \ref{tab:burst_parameters_2}.

Across all seven observing epochs, the intrinsic and scattering widths of the 74 bursts from this repeater range from $0.68$ to $29.53$ ms and $0.03$ to $14.47$ ms, respectively (Table~\ref{tab:literature_P1}). 
Since both intrinsic and scattering widths depend on observing frequency, we only consider  detections from this repeater reported in the literature within frequency ranges similar to our observing band (\textit{i.e.}, $250-500$ MHz). These include CHIME ($400-800$ MHz) \citep{amiri20,pearlman20,sand22}, GBT ($300-400$ MHz) \citep{chawla20}, SRT ($300-400$ MHz) \citep{trudu23}, and previous uGMRT observations in the $300-500$ MHz range \citep{sand22,trudu23}.
The reported intrinsic widths from these observations range from $0.58$ to $23.0$ ms, which almost overlap with the intrinsic widths observed in our study, where we detected some broadened pulses from $\FRB$. 
In contrast, only \citet{chawla20} and \citet{sand22} have reported the scattering width values for bursts from this repeater using GBT ($300-400$ MHz) and previous uGMRT ($300-500$ MHz) observations. Their reported values, ranging from $1.67$ to $5.90$ ms (Table~\ref{tab:literature_P1}), partially overlap with our observations, where we detected more scattered pulses from $\FRB$.

The spectrum of the burst has been fitted with a Gaussian function, expressed as 
\begin{equation}
I_{S}(f) = A_S \exp\left(-\frac{(f-\mu_S)^2}{2\sigma_S^2}\right)
\label{eq:spectra_fit}
\end{equation}
where, $I_S(f)$ is intensity of the spectra for a given frequency $f$, and the fitting parameters $A_S$, $\mu_S$ and $\sigma_S$ are determined by the least square method.
We have estimated the centre frequency and band occupancy of a burst using Eq.~\ref{eq:spectra_fit}.  
The parameter $\mu_S$ gives us the centre frequency ($\nu_{centre}$) of the burst which is a frequency value where the intensity of the spectra is maximum. 
The parameter $\sigma_S\times 2\ln 2$ gives the band occupancy ($\nu_{band}$) of the burst for which the intensity is dropped by $50\%$ of its peak value.  
The blue dashed line in the right panel of Figure~\ref{fig:sample_burst} shows the fitting of the spectra using eq.~\ref{eq:spectra_fit}. 
The values of centre frequency of the detected bursts for each of the observation epochs are tabulated in Tables~\ref{tab:burst_parameters_1} and \ref{tab:burst_parameters_2}.

Across all seven observing epochs, the measured values of $\nu_{centre}$ and $\nu_{band}$ were found to range $311.84-475.53$ MHz and $48.75-121.88$ MHz, respectively. However, no such information is available in the existing literature related to $\FRB$.


\section{Results}\label{sec:3}
In this section, we explore the variation of the physical parameter with the activity phase of $\FRB$ depending on the $74$ bursts detected from this repeater using Band-3 of uGMRT. 
We estimated the median values of the physical parameters and activity phases of this repeater across different observation epochs, with the goal of improving the modeling of the source mechanism and the local environment of the FRB.  
\subsection{Intrinsic Fluence}\label{sec:3.1}
We estimated the flux density ($S_{\nu}$) of each of the bursts detected at Band-3 of uGMRT using the calibrator $3C48$ \citep{ott1994}, which was observed along with $\FRB$ at each observation epoch.
The intrinsic fluence ($F_{int}$) of this repeater (represents the intrinsic energy emitted by the source) by multiplying the flux density ($S_{\nu}$) of each bursts with its intrinsic width ($w_{int}$) getting from the profile fitting discussed in the previous section. 
We considered here the median value of the intrinsic fluence, denoted as $\overline{F}_{int}$, and study its variation with $\phase$.

\begin{figure}
\centering
\includegraphics[width=\columnwidth]{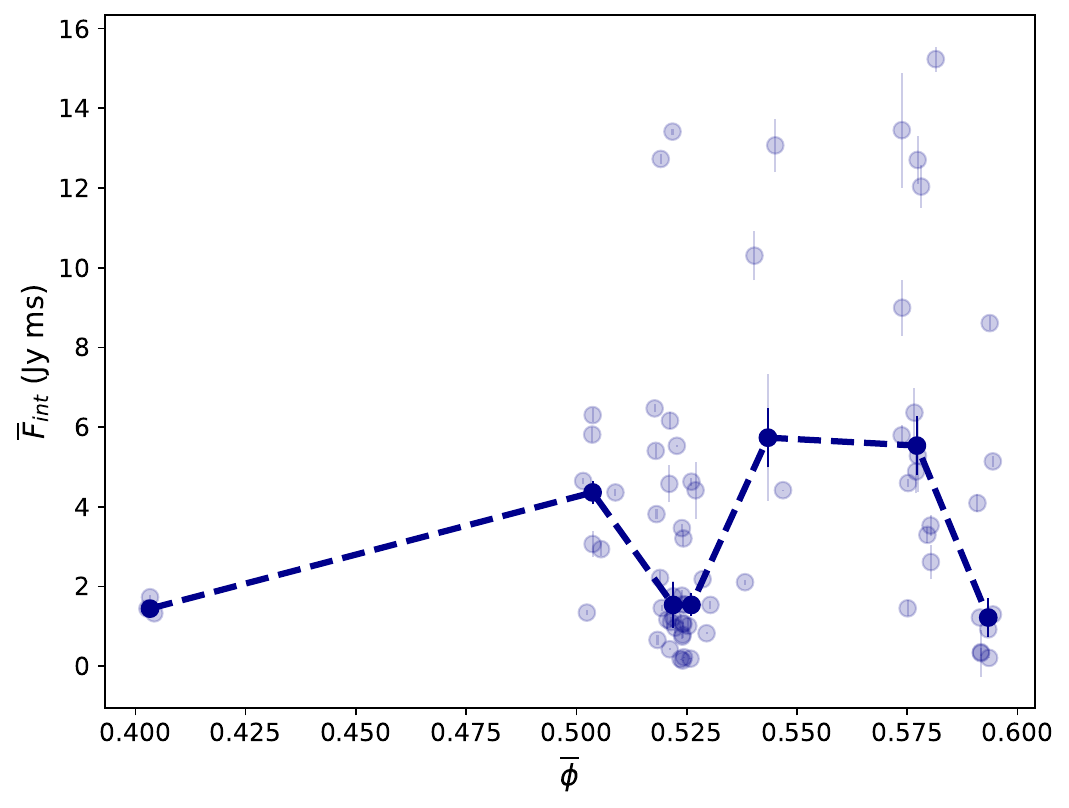}
\caption{The intrinsic fluence of all 74 bursts, along with the variation of their median values ($\overline{F}_{int}$), are shown as a function of the median activity phase ($\phase$) of the $\FRB$ at each observation epoch. The error bars represent the uncertainties in the estimates of $\overline{F}_{int}$, and $\phase$.}
\label{fig:phase_flux_fluence}
\end{figure}

The dark points connected by dashed lines in Figure~\ref{fig:phase_flux_fluence} show the variation of $\overline{F}_{int}$ with $\phase$ for this repeater, while the faint points represent the individual values of $F_{int}$ for all 74 bursts observed from it.  
The values of intrinsic fluence of the $74$ detected bursts for each of the observation epochs are tabulated in Tables~\ref{tab:burst_parameters_1} and \ref{tab:burst_parameters_2}.

Across all seven observing epochs, the intrinsic fluence of the 74 bursts from this repeater ranges from $0.14$ to $15.23$ Jy ms (Table~\ref{tab:literature_P2}). 
The intrinsic fluence depends on the observing frequency, so we consider only earlier detections with observing frequencies comparable to our observing band (\textit{i.e.}, $250-500$ MHz). The reported fluence from these studies ranges from $0.4$ to $172$ Jy ms (Table~\ref{tab:literature_P2}), with the intrinsic fluence values from our observations being a subset of this range, where we detected more bursts with lower intrinsic fluence from $\FRB$.

For each of the detected bursts, we estimate the value of $F_{int}$ from the corresponding value of $S_{\nu}$ using the relation $F_{int} = w_{int}\times S_{\nu}$, where $w_{int}$ has been estimated separately using profile fitting and no dependency on the value of $S_{\nu}$. 
However, the variation of median values of intrinsic width ($\overline{w}_{int}$) with $\phase$, which is not shown in this paper, has similar characteristics as we have seen in the variation of $\overline{F}_{int}$ with $\phase$.


\subsection{Excess $DM$ and Scattering Width}\label{sec:3.2}
The value of $DM$ gives us the integral electron column density whereas the value of $\tau$ gives us the nature of turbulence along the line-of-sight of the observation. 
Several studies (e.g. \citealt{bhat04}) find the correlation between the $DM$ and $\tau$ for Galactic pulsars.
However, no such correlation has been found between $DM$ and $\tau$ for extragalactic pulses such as FRBs.

Furthermore, the values of both $DM$ and $\tau$ are strongly influenced by burst morphology \citep{nimmo21,sand2023}, which is often complex in the case of FRBs. For example, the accuracy of estimating $DM$ and $\tau$ is closely tied to the time resolution and the amount of temporal substructure present in the bursts. Consequently, bursts with varying time resolutions or intricate temporal features can introduce significant uncertainties in the estimation of these parameters.
Additionally, if a burst inherently exhibits asymmetric or structured morphology, accurately measuring scattering timescales becomes challenging \citep{ocker2022}, as it requires disentangling intrinsic burst features from propagation effects such as scattering in the interstellar medium (ISM).
However, these sources of uncertainty are minimal in our analysis as all bursts were recorded with same time resolution, and they nearly all display simple, single-component, structureless profiles.

\begin{figure}
\centering
\includegraphics[width=\columnwidth]{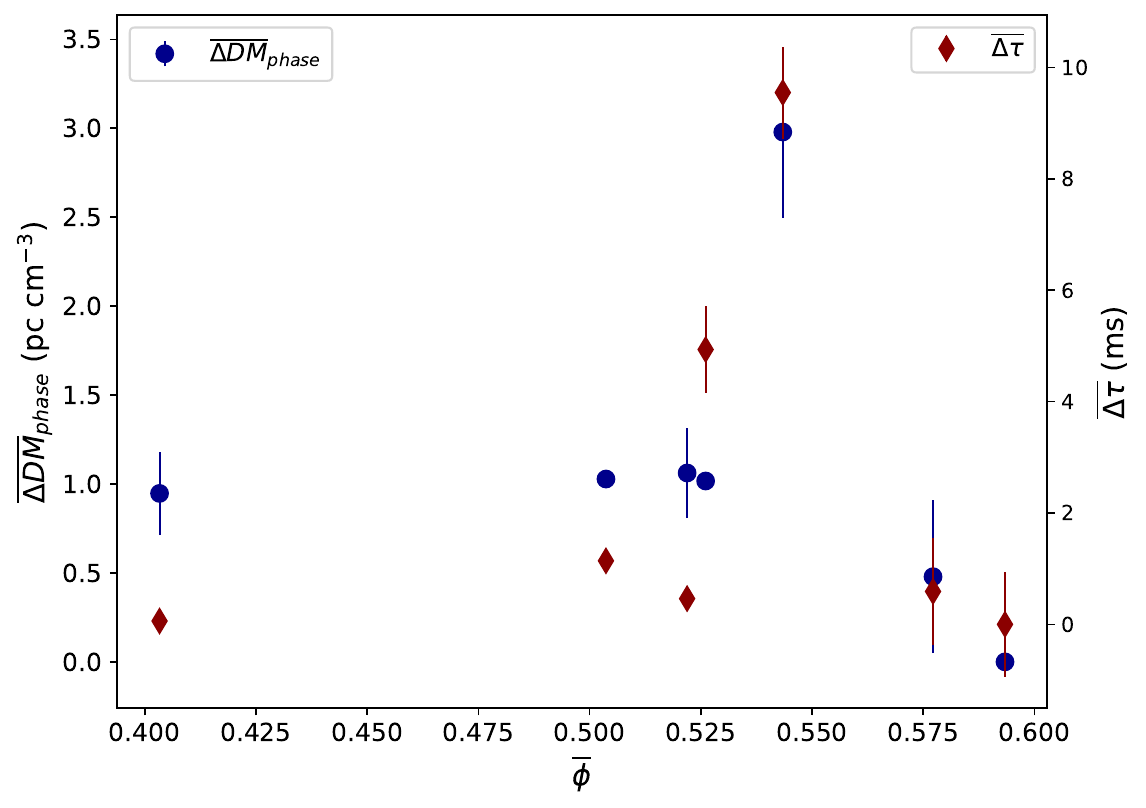}
\caption{The variation of median values of excess phase optimized $DM$ ($\overline{\Delta DM}_{phase}$) and excess scattering width ($\overline{\Delta \tau}$) with the median value of activity phases ($\phase$) of $\FRB$ at each of the observation epochs. Here the error bars show the uncertainty in the estimation of $\overline{\Delta DM}_{phase}$, $\overline{\Delta \tau}$ and $\phase$.}
\label{fig:phase_DM_scat}
\end{figure}

We now consider the excess value of $\overline{DM}_{phase}$ (discussed in Section~\ref{sec:2.4}), defined as $\overline{\Delta DM}_{phase} = \overline{DM}_{phase} - \overline{DM}_{phase}(\overline{\phi} = 0.6)$. 
Similarly, we consider the median value of the scattering width per scan, denoted by $\overline{\tau}$, and define the excess value of $\overline{\tau}$ as $\overline{\Delta \tau} = \overline{\tau} - \overline{\tau}(\overline{\phi} = 0.6)$. 
In other words, we study the excess values of both $\overline{DM}_{phase}$ and $\overline{\tau}$ relative to their minimum values observed at $\overline{\phi} = 0.6$. 
The values of $\overline{\Delta DM}_{phase}$ and $\Delta \tau$ are expected to exhibit evolution intrinsic to the source and/or its local environment. Figure~\ref{fig:phase_DM_scat} shows the variation of $\overline{\Delta DM}_{phase}$ and $\overline{\Delta \tau}$ with $\phase$.

In literature, Figure 5 of \citet{sand2023} shows no clear correlation between the variations of either $DM$ or $\tau$ with $\phase$. Moreover, there is no consistent qualitative trend observed in the variation of $DM$ or $\tau$ with $\phase$ individually. The ranges of $DM$ ($348.78-350.27$ pc cm$^{-3}$) and $\tau$ ($0.08-2.23$ ms) reported by \citet{sand2023} are narrower compared to the ranges of these parameters found in our analysis. 

\subsection{Centre Frequency and Band Occupancy}\label{sec:3.3}
$\FRB$ have been observed across a wide range of radio frequencies from  $110$ MHz to $8$ GHz (Table~\ref{tab:literature_P2}). 
The centre frequency ($\nu_{centre}$) of a burst observed detected in the uGMRT observation gives us the frequency at which its energy maximises, where the intrinsic frequency is different by a constant factor of (1+z).

Further the detected bursts show a ``band-limited'' structure, where the emission is strong in a limited frequency range, named as emission bandwidth ($\nu_{band}$) of the burst. 
In section-\ref{sec:2.5} we discussed how we arrived at the estimated value of $\nu_{centre}$ and $\nu_{band}$ using the spectra fitting of a burst (following Eq.~\ref{eq:spectra_fit}).
Here we discuss the variation of the median values of $\nu_{centre}$ per scan (denoted as $\overline{\nu}_{centre}$) and $\nu_{band}$ (denoted as $\overline{\nu}_{band}$), with $\phase$ of $\FRB$ at each of the observation epoch.

\begin{figure}
\centering
\includegraphics[width=\columnwidth]{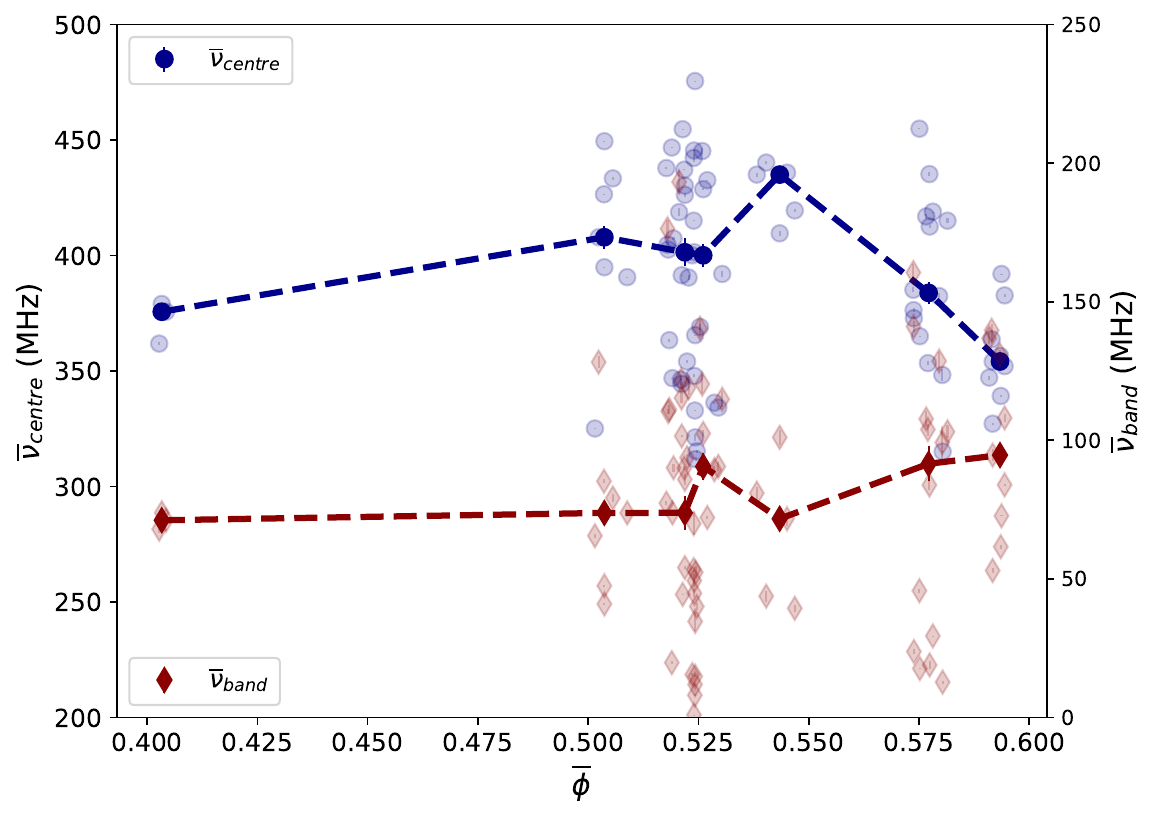}
\caption{The centre frequency and band occupancy of all 74 bursts, along with the variation of their median values ($\overline{\nu}_{centre}$ and $\overline{\nu}_{band}$), are shown as a function of the median activity phase ($\phase$) of the $\FRB$ at each observation epoch. The error bars represent the uncertainties in the estimates of $\overline{\nu}_{centre}$, $\overline{\nu}_{band}$, and $\phase$.}
\label{fig:phase_freq_band}
\end{figure}

The dark points connected by dashed lines in Figure~\ref{fig:phase_freq_band} show the variation of $\overline{\nu}_{centre}$ and $\overline{\nu}_{band}$ with $\phase$ for this repeater, while the faint points represent the individual values of $\nu_{centre}$ and $\nu_{band}$ for all 74 bursts observed from it.

\citet{bethapudi2023} reported a chromatic dependence of $\nu_{centre}$ on activity phase, based on observations of this repeater across multiple telescopes spanning a wide frequency range (100 MHz to 10 GHz). In this context, we suggest that the variation in $\overline{\nu}_{\text{centre}}$ observed in our study is likely a local effect, influenced by the narrower frequency coverage (250$–$500 MHz) and the smaller sample of bursts compared to previous reports in the literature.

\citet{sand2023} reported that $\nu_{band}$ is largely appeared to be scattered. The variation of $\overline{\nu}_{band}$ observed in our analysis may reflect a local behaviour of this parameter.

\subsection{Waiting Time}\label{sec:3.4}
The arrival time of a burst is obtained from the single$\_$pulse$\_$search tool which gives the candidate time ($T_{\rm cand}$) by setting the clock to zero at the start of each scan of each observation epoch.
For each of the observation epochs, the burst waiting time ($T_{\rm wait}$) is estimated by considering the $T_{\rm cand}$ difference between two consecutive bursts. 
We have repeated this process for all other observation epochs separately.

\begin{figure}
\centering
\includegraphics[width=\columnwidth]{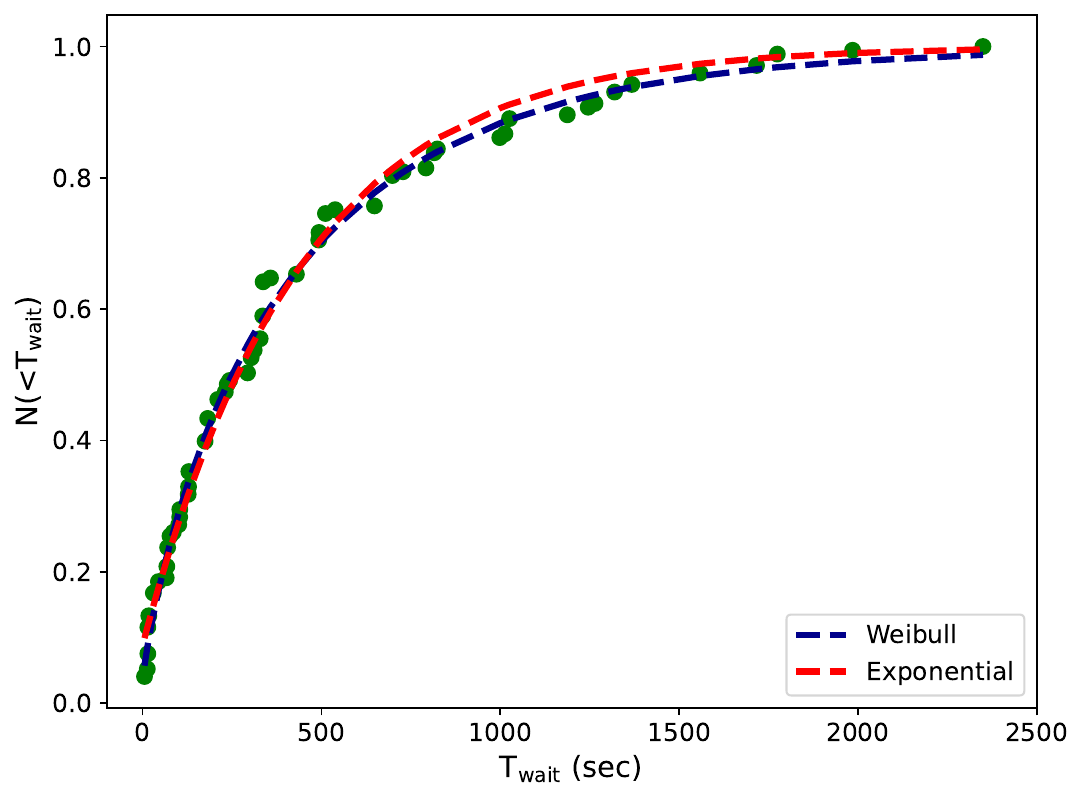}
\caption{The normalized cumulative distribution ($N(<T_{\rm wait})$) of the waiting time ($T_{\rm wait}$) between consecutive bursts. The blue and red dashed lines are the fitting with a Weibull and an exponent function respectively.}
\label{fig:waiting_time_cumulative}
\end{figure}

The study of waiting time distribution is crucial for understanding the physical nature of the source. The normalized cumulative burst rate with a waiting time $T_{\rm wait}$ is denoted as $N(<T_{\rm wait})$. 
The green points in Figure~\ref{fig:waiting_time_cumulative} illustrate the variation of $N(<T_{\rm wait})$ with $T_{\rm wait}$.
In many cases, researchers often fit the $N(<T_{\rm wait})$ vs. $T_{\rm wait}$ curve using a normalized Weibull distribution, defined as $W_{CDF}(\delta | k, r) = \exp[-(\delta r \Gamma(1 + 1/k))^k]$ \citep{oppermann18}, where the parameter $ k \to 0 $ converts the Weibull distribution into an exponential distribution. 
The blue dashed line in Figure~\ref{fig:waiting_time_cumulative} shows the Weibull distribution with coefficient $k = 0.84 \pm 0.02 $ and burst rate $r=8.01\pm 0.15$ hr$^{-1}$, indicating that the Weibull distribution is converging toward the exponential distribution.
To further investigate this convergence, we also fit the $N(<T_{\rm wait})$ vs. $T_{\rm wait}$ curve using a normalized exponential function, represented by the green dashed line in Figure~\ref{fig:waiting_time_cumulative}. 
The reduced $\chi^2$ values, which assess the goodness of fit, are unity for both the Weibull distribution (blue dashed line) and the normalized exponential function (red dashed line). 
We note that the arrival time of the bursts have been corrected using the Kaplan-Meier estimator \citep{kaplan1958}. 
In summary, the $N(<T_{\rm wait})$ vs. $T_{\rm wait}$ curve is well-fitted by an exponential function, indicating that the $74$ detected bursts from $\FRB$ follow Poissonian statistics, suggesting that these bursts are non-clustered.

\subsection{Burst Rate}\label{sec:3.5}
\begin{figure}
\centering
\includegraphics[width=\columnwidth]{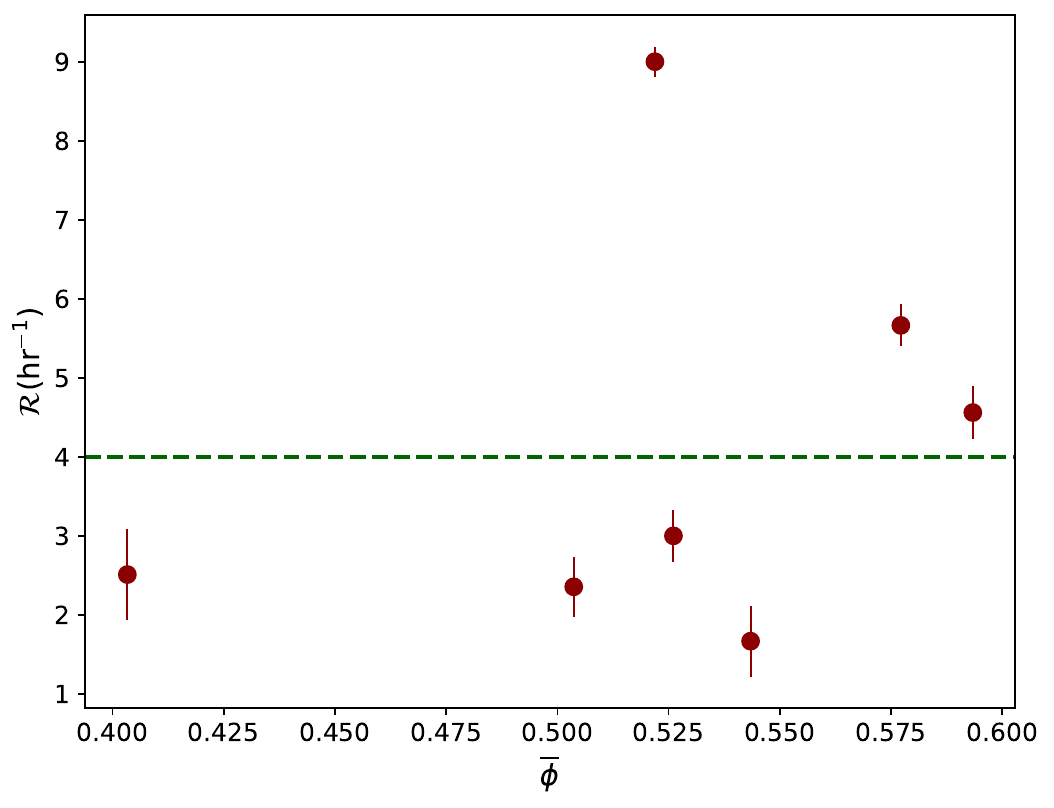}
\caption{The variation of burst rate ($\mathcal{R}$) with the median value of activity phases ($\phase$) of $\FRB$ at each of the observation epochs. Here the error bars show the $1\sigma$ uncertainty in the estimation of $\mathcal{R}$ and the green horizontal dashed line shows the burst rate averaged over all the observation epochs.}
\label{fig:phase_burst_rate}
\end{figure}

Here we discuss the burst rate of $\FRB$ depending on the $74$ bursts detected at Band-3 of uGMRT. 
This discussion breaks into two parts.
First, we discuss the variation of burst rate with phase and then the variation of cumulative burst rate with the fluence of the burst. 
We have estimated the burst rate, which is the number of bursts divided by the total observation duration, at each observation epoch separately and studied its variation with the median phase value for each epoch. 
Figure~\ref{fig:phase_burst_rate} shows the variation of burst rate ($\mathcal{R}$) with $\phase$ at each of the observation epochs.  
\citet{sand2023} showed that the burst rate of this repeater varies as a Gaussian function, based on over 200 hours of observations. In contrast, it is not meaningful to determine the maximum burst rate from our analysis, given the significantly shorter total exposure time of only 18 hours, which is statistically limited compared to \citet{sand2023}. In this context, we instead report the average burst rate of the $\FRB$, indicated by the green horizontal dashed line in Figure~\ref{fig:phase_burst_rate}, calculated by combining all observation epochs. We find that the average burst rate is approximately 4 bursts per hour at an observing frequency of 400 MHz, for bursts with fluence above 0.14 Jy ms.
This average burst rate is higher for this repeater reported by \citet{chawla20}, \citet{pleunis21}, and \citet{sand22} (Table~\ref{tab:literature_P2}), which are $1.8$ bursts per hour with fluence above $5$ Jy ms at $600$ MHz, $0.3$ bursts per hour with fluence above $26$ Jy ms at $150$ MHz, and $0.7$ bursts per hour with fluence above $0.5$ Jy ms at $400$ MHz (from an earlier uGMRT observation), respectively. 
The burst rate predicted from our observation is comparable to the rate reported by \citet{sand22} (Table~\ref{tab:literature_P2}), which is $4.2$ bursts per hour with fluence above $0.2$ Jy ms at $800$ MHz. However, it is smaller than the burst rate reported by \citet{bethapudi23} (Table~\ref{tab:literature_P2}), which is $22.8$ bursts per hour with fluence above $0.005$ Jy ms at $6$ GHz. 
We note that the Euclidean scaling from a fluence of 0.14 Jy ms to 0.005 Jy ms results in an increase in the burst rate from 4 to 18 bursts per hour, which is comparable to the burst rate reported by \citet{bethapudi23}.
It should also be noted that burst rate strongly depends on the observing frequency and the threshold fluence value of the telescope.

We now discuss the variation of the cumulative burst rate $\mathcal{R}_{\rm cum}(>F_{\rm int})$ with the intrinsic fluence $F_{\rm int}$ of the bursts from $\FRB$. 
This is defined as the number of bursts having an intrinsic fluence greater than $F_{\rm int}$, divided by the observation time. We assessed the fluence completeness of all detected bursts using the method described in \citet{keane15}.

\begin{figure}
\centering
\includegraphics[width=\columnwidth]{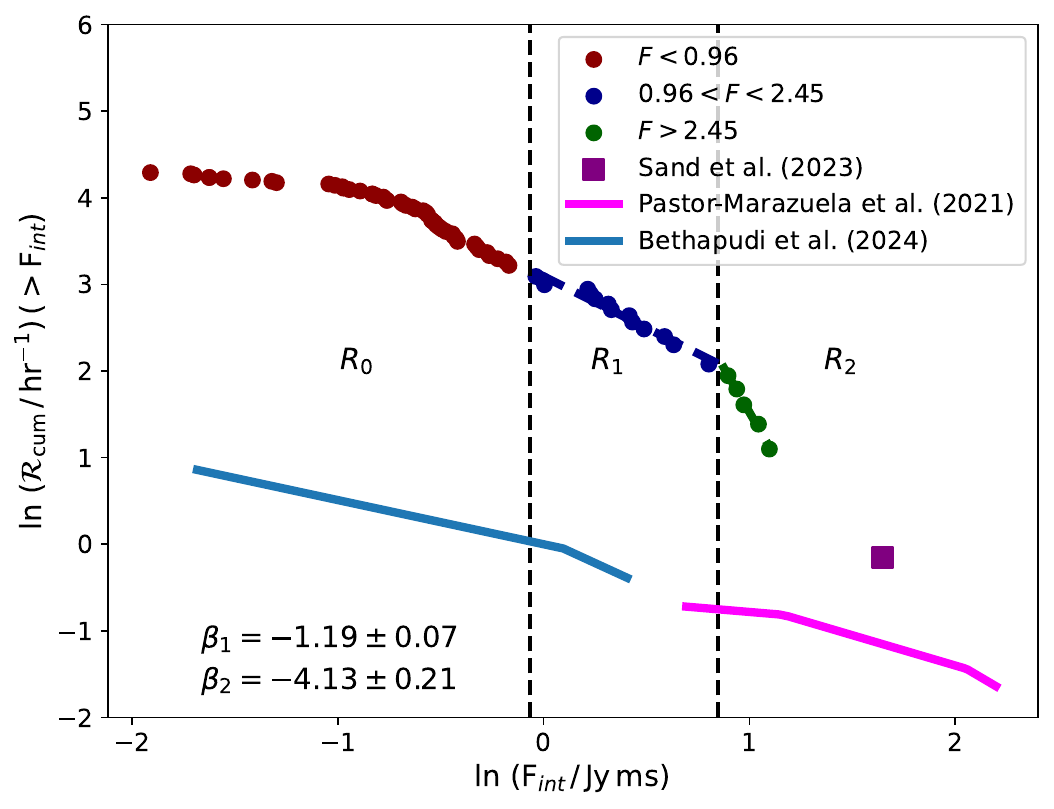}
\caption{The variation of the cumulative burst rate ($\mathcal{R}_{\rm cum}$) with the intrinsic fluence (${\rm F}_{\rm int}$) emitted by the source.}
\label{fig:burst_rate_energy_all}
\end{figure}

Figure~\ref{fig:burst_rate_energy_all} shows the variation of the cumulative burst rate ($\mathcal{R}_{\rm cum}(>F_{\rm int})$) with the intrinsic fluence ($F_{\rm int}$) for all $74$ bursts from $\FRB$. 
The variation of $\mathcal{R}_{\rm cum}(>F_{\rm int})$ vs. $F_{\rm int}$ is divided into three regions, namely $R_0$, $R_1$, and $R_2$. 
In region $R_0$, which includes bursts from this repeater with an intrinsic fluence value $F_{\rm int} < 0.96$ Jy ms, all bursts are fluence incomplete and are therefore not considered for further modelling. 
In contrast, regions $R_1$ ($0.96$ Jy ms $< F_{\rm int} < 2.45$ Jy ms) and $R_2$ ($F_{\rm int} > 2.45$ Jy ms) contain fluence-complete bursts, which were used to model the $\mathcal{R}_{\rm cum}(>F_{\rm int})$ vs. $F_{\rm int}$ variation.
We found that the fluence-complete bursts from $\FRB$ can be fitted using a broken power law with indices $\beta_1 = -1.19 \pm 0.07$ for region $R_1$ and $\beta_2 = -4.13 \pm 0.21$ for region $R_2$, with the break in intrinsic fluence located at $2.45$ Jy ms. 
The reduced $\chi^2$ value (which determines the goodness of fit) for this fitting is $0.98$ (close to unity), and the fitted curves for both $R_1$ and $R_2$ are shown as dashed lines in Figure~\ref{fig:burst_rate_energy_all}. 

Previously, \citet{amiri20} reported a power law modelling of the cumulative burst rate for $\FRB$ with a single power law index of $-3.3$. 
However, they did not identify any turnover in the distribution at lower fluences, likely due to the sensitivity limit of CHIME ($400-800$ MHz) or the intrinsic burst distribution. 
Later, \citet{pastor2021} reported power law modelling of the cumulative burst rate for this repeater using the Apertif ($1220-1520$ MHz) telescope, identifying two turnovers. 
For $F > 7.8$ Jy ms, the power law index was $-1.4 \pm 0.1$; for $3.2$ Jy ms $ < F < 7.8$ Jy ms, the index was $-0.7 \pm 0.1$; and for $F < 3.2$ Jy ms, the index was $-0.2 \pm 0.1$.
The studies mentioned above mainly probe the high fluence bursts from this repeater observed at relatively high frequencies. 
In contrast, our observations detected low intrinsic fluence bursts from $\FRB$ at a low-frequency regime, showing a broken power law in its cumulative burst rate with power law indices that change sharply with the turnover of intrinsic fluence. Recently, \citet{bethapudi24} identified two fluence turnovers at $0.18$ Jy ms and $1.10$ Jy ms using $79$ bursts from $\FRB$ at uGMRT (550–750 MHz), with $0.18$ Jy ms considered as the fluence completeness limit. 
The reported power indices \citep{bethapudi24} are $-0.51 \pm 0.01$ for $F < 1.10$ Jy ms and $-1.09 \pm 0.07$ for $F > 1.10$ Jy ms. However, they reported bursts with lower fluence values and relatively flatter power indices compared to our observations.
\citet{sand2023} reported the burst rate of $\FRB$ is $0.86^{+0.5}_{-0.35}$ burst per hour at a fluence threshold of 5.2 Jy ms centred around the peak activity of the source ($0.45-0.55$).

Along with our results for this repeater, Figure~\ref{fig:burst_rate_energy_all} also shows the cumulative burst rate, as a function of intrinsic fluence reported by \citet{pastor2021} (magenta line), \citet{bethapudi24} (navy line), and \citet{sand2023} (purple point). 
The cumulative burst rates of $\FRB$ mentioned above (Figure~\ref{fig:burst_rate_energy_all}) are measured at different frequencies: $120$ MHz \citep{pastor2021}, $650$ MHz \citep{bethapudi24}, $600$ MHz \citep{sand2023}, and $375$ MHz (current analysis). To enable a consistent comparison of the cumulative burst rates of this repeater across these frequencies, we scale them to our observational frequency ($375$ MHz) using equation~2 of \citet{houben19}.
This comparison indicates that the burst rate inferred from our analysis is higher than that reported in previous observations. Such differences may arise from several factors, including possible temporal evolution of the source activity and the use of different observing frequencies and instrumental sensitivities across different telescopes.
Further, other repeaters as well as giant pulse emitters also show such varying burst rates.


\subsection{Multi-Component Burst Timescales}\label{sec:3.6}
\begin{figure}
\centering
\includegraphics[width=\columnwidth]{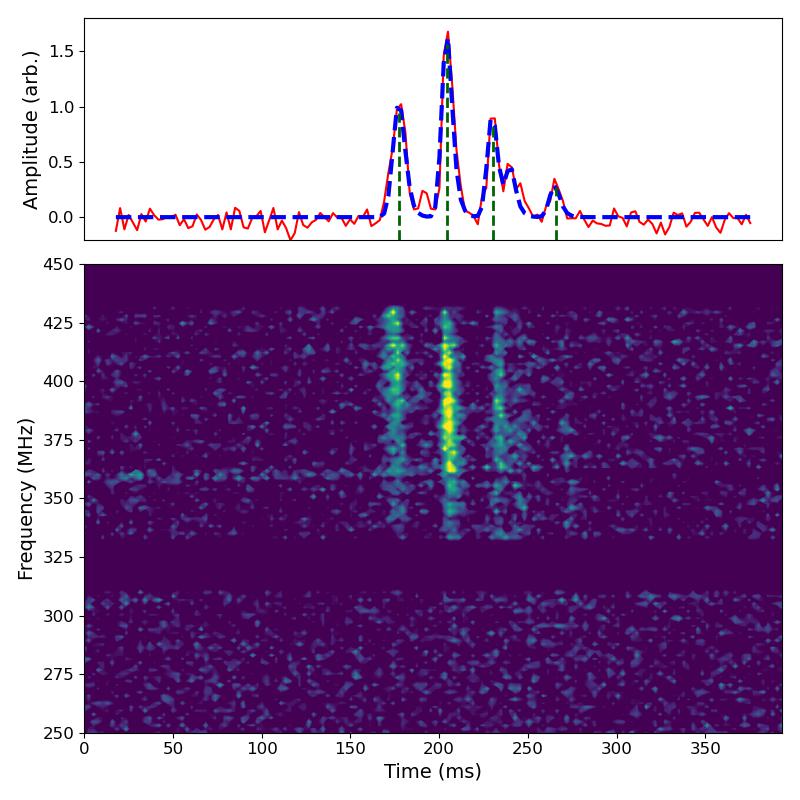}
\includegraphics[width=\columnwidth]{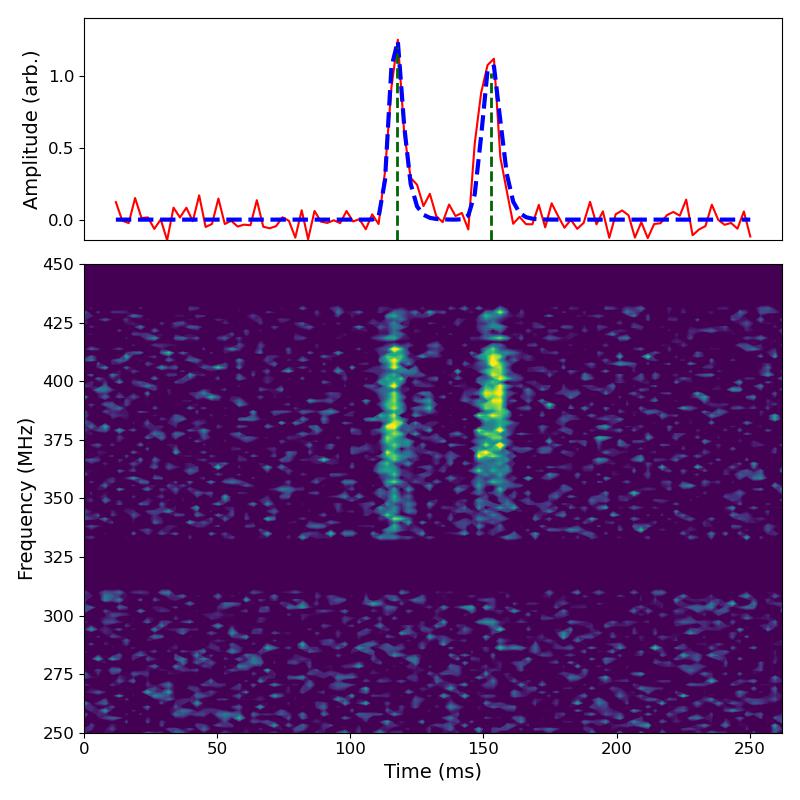}
\caption{The dynamic spectra and profiles of two multi-component bursts from $\FRB$ at activity phase $\phi = 0.527$.}
\label{fig:triple_burst}
\end{figure}

We detected $74$ bursts in Band-3, the majority of which are single-component and weakly structured. Only two bursts exhibit clear multi-component sub-structure.
Figure~\ref{fig:triple_burst} shows these two events observed at activity phase $\phi = 0.527$, where one with four components separated by $29.62$ ms, $28.05$ ms, and $38.79$ ms, and another with two components separated by $37.5$ ms.

The observed separations ($28-39$ ms) are therefore more likely to reflect stochastic sub-structure or closely clustered emission rather than a coherent periodic signal. A significantly larger sample of multi-component bursts will be required to explore such connections reliably.

\subsection{Frequency Drifting} \label{sec:3.7}
The time–frequency downward drifting pattern is a common characteristic of repeaters \citep{wang2019}. \citet{chawla20, pearlman20, sand22, trudu23} reported linear drift rates for the bursts from $\FRB$ as $-4.2$ MHz ms$^{-1}$, $-6.6$ MHz ms$^{-1}$, $-8.0$ MHz ms$^{-1}$, $-9.7$ MHz ms$^{-1}$, and $-5.9$ MHz ms$^{-1}$, using observations from GBT ($300-400$ MHz), LOFAR ($110-188$ MHz), GBT ($600-1000$ MHz), SRT ($300-400$ MHz), and uGMRT ($300-500$ MHz), respectively. 

In contrast, we did not observe any linear drifting pattern in the dynamic spectra for the 74 bursts from this repeater, while most of the bursts detected in our observation are single-component in nature.

\section{Summary and Discussion}\label{sec:4}
We conducted an interferometric observation of $\FRB$ from November 26, 2020, to March 21, 2021, over seven different epochs, sampling the activity phase space of this repeater from $\phi = 0.4$ to $\phi = 0.6$. Simultaneous phased array observations were performed at Band-3 ($250-500$ MHz), Band-4 ($550-850$ MHz), and Band-5 ($1060-1460$ MHz) of uGMRT, covering a wide frequency range of $250-1460$ MHz with a frequency resolution of approximately $48.83$ kHz and a time resolution of $655.36$ $\mu$s.
In our observations, we detected 74 bursts at Band-3 and 4 bursts at Band-4, while no bursts were detected at Band-5. Additionally, we did not find any simultaneous detections of bursts at both Band-3 and Band-4. In this paper, we focus solely on the 74 bursts detected at Band-3 to examine the variations in their physical parameters. 
We discuss the variations in the observed parameters of these 74 bursts with the activity phase of $\FRB$ and compare our findings with existing literature on this repeater. The qualitative and quantitative nature of these physical parameter variations provides valuable insights for theoretical astrophysicists, helping to enhance the modeling of the source mechanism and the local environment of the $\FRB$.\\

Using this sample of 74 bursts in band-3 (250$-$500 MHz) and 4 bursts in band-4 (550$-$850 MHz), it was observed that the median centre frequency, along with the excess median dispersion measure and scattering, all reach their maximum values around the middle of the activity phase. This possibly indicates that the bursts encounter denser plasma closer to the centre of the activity window. This suggests an orbital motion (e.g. \citealt{lyutikov2020}), with the activity window centred around the point of closest approach possibly involving interactions with a companion star or a stellar wind. The excess median dispersion measure at this point constrains the plasma density within the orbit, while the variation in the median centre frequency likely provides insights into the emission process itself. Gaining a deeper understanding of its progenitor could offer valuable insights into the broader origins of FRBs.

\section*{Acknowledgement}
We acknowledge the support of the Department of Atomic Energy, Government of India, under project no. 12-R\&D-TFR-5.02-0700. 
The GMRT is run by the National Centre for Radio Astrophysics of the Tata Institute of Fundamental Research, India. 
We acknowledge the support of GMRT telescope operators for observations. 
We appreciate the insightful inputs provided by Ujjwal Panda and Chahat Dudeja during FRB data analysis.

\begin{deluxetable*}{lcccccccccc}
\tablenum{1}
\label{tab:burst_parameters_1}
\tablecaption{Observed and derived parameters of the 74 bursts detected from $\FRB$ at Band-3 of uGMRT. The symbols $DM_{phase}$, $w_{int}$, $\tau$, $S_{\nu}$, $F_{int}$ and $\nu_{centre}$ stand for dispersion measure (phase optimized), intrinsic width, scattering width, flux density, intrinsic fluence and centre frequency respectively.}
\tablewidth{\textwidth}
\tabletypesize{\scriptsize}
\tablehead{
\colhead{Epoch} & 
\colhead{Duration} & 
\colhead{No. of} & 
\colhead{Burst} & 
\colhead{Phase} & 
\colhead{$DM_{phase} \pm \delta DM_{phase}$} & 
\colhead{$w_{int} \pm \delta w_{int}$} & 
\colhead{$\tau \pm \delta \tau$} & 
\colhead{$S_\nu$} & 
\colhead{$F_{int} \pm \delta F_{int}$} & 
\colhead{$\nu_{centre} \pm \delta \nu_{centre}$} \\
\colhead{} & 
\colhead{(hour)} & 
\colhead{bursts} & 
\colhead{ID} & 
\colhead{} & 
\colhead{(pc cm$^{-3}$)} & 
\colhead{(ms)} & 
\colhead{(ms)} & 
\colhead{(Jy)} & 
\colhead{(Jy ms)} & 
\colhead{(MHz)} }
\startdata
21 Mar 2021 & $2.0$ & $9$ & 01 & $0.591000$ & $349.02 \pm 0.13$ & $15.10 \pm 2.45$ & $0.40 \pm 0.01$ & $0.27$ & $4.09 \pm 0.67$ & $347.15 \pm 3.49$ \\
 &  &  & 02 & $0.591000$ & $348.94 \pm 0.13$ & $7.45 \pm 1.68$ & $0.22 \pm 0.01$ & $0.16$ & $1.22 \pm 0.27$ & $363.72 \pm 3.82$ \\
 &  &  & 03 & $0.592000$ & $351.90 \pm 0.11$ & $1.01 \pm 0.90$ & $0.52 \pm 0.46$ & $0.35$ & $0.35 \pm 0.31$ & $327.22 \pm 1.77$ \\
 &  &  & 04 & $0.592000$ & $350.17 \pm 0.13$ & $1.54 \pm 8.32$ & $8.33 \pm 2.85$ & $0.21$ & $0.33 \pm 1.76$ & $354.13 \pm 3.55$ \\
 &  &  & 05 & $0.593000$ & $349.04 \pm 0.13$ & $4.45 \pm 1.94$ & $6.43 \pm 2.13$ & $0.21$ & $0.92 \pm 0.40$ & $356.56 \pm 2.95$ \\
 &  &  & 06 & $0.594000$ & $348.94 \pm 0.20$ & $6.62 \pm 0.52$ & $6.29 \pm 0.51$ & $0.78$ & $5.14 \pm 0.40$ & $352.11 \pm 0.71$ \\
 &  &  & 07 & $0.594000$ & $349.01 \pm 0.10$ & $6.09 \pm 0.94$ & $0.22 \pm 0.01$ & $0.21$ & $1.30 \pm 0.20$ & $382.74 \pm 2.28$ \\
 &  &  & 08 & $0.594000$ & $348.98 \pm 0.14$ & $0.72 \pm 0.12$ & $17.20 \pm 3.23$ & $0.29$ & $0.21 \pm 0.04$ & $339.20 \pm 1.32$ \\
 &  &  & 09 & $0.594000$ & $349.02 \pm 0.10$ & $3.59 \pm 0.22$ & $2.45 \pm 0.16$ & $2.40$ & $8.61 \pm 0.52$ & $392.02 \pm 0.30$ \\
\hline
20 Mar 2021 & $3.0$ & $27$ & 10 & $0.518000$ & $349.87 \pm 0.13$ & $28.34 \pm 2.81$ & $0.54 \pm 0.01$ & $0.19$ & $5.40 \pm 0.54$ & $404.68 \pm 5.00$ \\
 &  &  & 11 & $0.518000$ & $349.02 \pm 0.11$ & $4.27 \pm 0.43$ & $2.22 \pm 0.35$ & $0.89$ & $3.81 \pm 0.38$ & $402.57 \pm 1.27$ \\
 &  &  & 12 & $0.518000$ & $345.30 \pm 0.12$ & $4.45 \pm 2.64$ & $0.45 \pm 0.09$ & $0.15$ & $0.66 \pm 0.39$ & $363.35 \pm 2.76$ \\
 &  &  & 13 & $0.519000$ & $348.91 \pm 0.15$ & $14.56 \pm 3.27$ & $0.56 \pm 0.01$ & $0.15$ & $2.21 \pm 0.50$ & $446.74 \pm 1.09$ \\
 &  &  & 14 & $0.519000$ & $349.93 \pm 0.13$ & $23.69 \pm 0.66$ & $0.21 \pm 0.01$ & $0.54$ & $12.73 \pm 0.35$ & $346.89 \pm 0.76$ \\
 &  &  & 15 & $0.519000$ & $349.02 \pm 0.07$ & $3.82 \pm 0.55$ & $0.12 \pm 0.01$ & $0.38$ & $1.46 \pm 0.21$ & $407.37 \pm 2.50$ \\
 &  &  & 16 & $0.518000$ & $349.02 \pm 0.07$ & $5.74 \pm 0.27$ & $0.41 \pm 0.01$ & $1.13$ & $6.47 \pm 0.30$ & $437.84 \pm 1.23$ \\
 &  &  & 17 & $0.521000$ & $348.46 \pm 0.20$ & $5.01 \pm 2.96$ & $1.14 \pm 3.33$ & $0.23$ & $1.17 \pm 0.69$ & $418.78 \pm 6.77$ \\
 &  &  & 18 & $0.521000$ & $349.02 \pm 0.10$ & $1.03 \pm 0.37$ & $1.98 \pm 0.56$ & $1.07$ & $1.11 \pm 0.39$ & $454.68 \pm 0.43$ \\
 &  &  & 19 & $0.522000$ & $349.36 \pm 0.19$ & $11.78 \pm 0.20$ & $1.17 \pm 0.12$ & $1.14$ & $13.41 \pm 0.23$ & $437.15 \pm 1.13$ \\
 &  &  & 20 & $0.522000$ & $348.93 \pm 0.13$ & $6.58 \pm 1.03$ & $0.09 \pm 0.01$ & $0.18$ & $1.22 \pm 0.19$ & $426.28 \pm 3.20$ \\
 &  &  & 21 & $0.522000$ & $349.01 \pm 0.05$ & $3.90 \pm 2.62$ & $0.63 \pm 0.32$ & $0.45$ & $1.75 \pm 1.18$ & $430.25 \pm 0.80$ \\
 &  &  & 22 & $0.522000$ & $351.83 \pm 0.22$ & $6.46 \pm 2.43$ & $0.20 \pm 0.01$ & $0.15$ & $0.96 \pm 0.36$ & $354.11 \pm 2.30$ \\
 &  &  & 23 & $0.523000$ & $349.08 \pm 0.13$ & $21.43 \pm 0.21$ & $1.62 \pm 0.04$ & $0.26$ & $5.53 \pm 0.05$ & $390.47 \pm 4.13$ \\
 &  &  & 24 & $0.521000$ & $350.25 \pm 0.14$ & $13.19 \pm 4.03$ & $3.45 \pm 4.46$ & $0.35$ & $4.58 \pm 1.40$ & $346.30 \pm 2.79$ \\
 &  &  & 25 & $0.521000$ & $350.86 \pm 0.17$ & $1.99 \pm 0.32$ & $0.16 \pm 0.05$ & $0.21$ & $0.42 \pm 0.07$ & $344.36 \pm 3.07$ \\
 &  &  & 26 & $0.521000$ & $349.04 \pm 0.15$ & $5.27 \pm 0.53$ & $0.85 \pm 0.08$ & $1.17$ & $6.16 \pm 0.62$ & $391.47 \pm 0.80$ \\
 &  &  & 27 & $0.525000$ & $348.98 \pm 0.08$ & $3.88 \pm 1.18$ & $2.82 \pm 1.00$ & $0.40$ & $1.54 \pm 0.47$ & $315.27 \pm 2.31$ \\
 &  &  & 28 & $0.524000$ & $351.93 \pm 0.15$ & $3.80 \pm 0.84$ & $0.03 \pm 0.01$ & $0.20$ & $0.74 \pm 0.16$ & $445.45 \pm 0.12$ \\
 &  &  & 29 & $0.524000$ & $350.57 \pm 0.14$ & $4.28 \pm 1.42$ & $0.04 \pm 0.01$ & $0.18$ & $0.79 \pm 0.26$ & $401.40 \pm 3.82$ \\
 &  &  & 30 & $0.524000$ & $346.74 \pm 0.17$ & $0.68 \pm 0.02$ & $3.20 \pm 1.53$ & $0.20$ & $0.14 \pm 0.01$ & $347.88 \pm 2.54$ \\
 &  &  & 31 & $0.524000$ & $350.38 \pm 0.16$ & $11.13 \pm 2.78$ & $0.55 \pm 0.01$ & $0.16$ & $1.77 \pm 0.44$ & $442.24 \pm 6.55$ \\
 &  &  & 32 & $0.524000$ & $352.87 \pm 0.14$ & $9.34 \pm 0.45$ & $4.47 \pm 2.58$ & $0.12$ & $1.08 \pm 0.05$ & $332.91 \pm 0.47$ \\
 &  &  & 33 & $0.524000$ & $349.71 \pm 0.15$ & $7.29 \pm 1.63$ & $0.37 \pm 0.01$ & $0.14$ & $1.05 \pm 0.23$ & $475.52 \pm 0.60$ \\
 &  &  & 34 & $0.524000$ & $349.05 \pm 0.14$ & $6.69 \pm 1.24$ & $4.61 \pm 1.18$ & $0.48$ & $3.20 \pm 0.59$ & $365.53 \pm 0.58$ \\
 &  &  & 35 & $0.524000$ & $349.28 \pm 0.15$ & $10.15 \pm 1.87$ & $0.40 \pm 0.01$ & $0.15$ & $1.55 \pm 0.29$ & $321.28 \pm 5.83$ \\
 &  &  & 36 & $0.524000$ & $356.71 \pm 0.18$ & $1.19 \pm 0.04$ & $3.20 \pm 2.87$ & $0.19$ & $0.22 \pm 0.01$ & $311.84 \pm 3.69$ \\
\hline
15 Feb 2021 & 3.0 & $7$ & 37 & $0.502000$ & $349.00 \pm 0.08$ & $4.07 \pm 0.66$ & $0.40 \pm 0.01$ & $0.33$ & $1.34 \pm 0.22$ & $407.90 \pm 3.38$ \\
 &  &  & 38 & $0.502000$ & $349.04 \pm 0.12$ & $13.75 \pm 0.60$ & $1.24 \pm 0.10$ & $0.34$ & $4.65 \pm 0.20$ & $325.11 \pm 1.34$ \\
 &  &  & 39 & $0.504000$ & $349.45 \pm 0.16$ & $9.69 \pm 0.97$ & $3.31 \pm 0.98$ & $0.60$ & $5.81 \pm 0.58$ & $426.50 \pm 1.18$ \\
 &  &  & 40 & $0.504000$ & $349.20 \pm 0.27$ & $10.10 \pm 3.29$ & $9.78 \pm 3.23$ & $0.30$ & $3.06 \pm 1.00$ & $394.92 \pm 0.74$ \\
 &  &  & 41 & $0.504000$ & $349.05 \pm 0.15$ & $5.73 \pm 0.49$ & $2.75 \pm 0.47$ & $1.10$ & $6.30 \pm 0.54$ & $449.48 \pm 0.40$ \\
 &  &  & 42 & $0.506000$ & $347.87 \pm 0.15$ & $14.25 \pm 2.45$ & $0.49 \pm 0.00$ & $0.21$ & $2.94 \pm 0.51$ & $433.39 \pm 2.18$ \\
 &  &  & 43 & $0.509000$ & $349.05 \pm 0.11$ & $5.27 \pm 0.32$ & $0.45 \pm 0.02$ & $0.83$ & $4.36 \pm 0.26$ & $390.63 \pm 0.71$ \\
\hline
13 Feb 2021 & $1.5$ & $3$ & 44 & $0.403000$ & $352.05 \pm 0.10$ & $6.99 \pm 2.37$ & $0.22 \pm 0.01$ & $0.21$ & $1.44 \pm 0.49$ & $361.86 \pm 2.35$ \\
 &  &  & 45 & $0.403000$ & $348.97 \pm 0.14$ & $4.55 \pm 0.44$ & $0.15 \pm 0.01$ & $0.38$ & $1.73 \pm 0.17$ & $379.04 \pm 1.05$ \\
 &  &  & 46 & $0.404000$ & $348.54 \pm 0.17$ & $7.80 \pm 1.47$ & $0.16 \pm 0.01$ & $0.17$ & $1.32 \pm 0.25$ & $375.66 \pm 2.58$ \\
\hline
28 Dec 2020 & $3.0$ & $9$ & 47 & $0.524000$ & $350.18 \pm 0.17$ & $1.17 \pm 0.16$ & $5.91 \pm 0.53$ & $0.15$ & $0.18 \pm 0.02$ & $400.09 \pm 0.88$ \\
 &  &  & 48 & $0.524000$ & $348.98 \pm 0.09$ & $8.63 \pm 0.75$ & $0.04 \pm 0.01$ & $0.40$ & $3.46 \pm 0.30$ & $415.09 \pm 1.34$ \\
 &  &  & 49 & $0.526000$ & $349.12 \pm 0.19$ & $0.83 \pm 0.05$ & $14.47 \pm 6.11$ & $0.23$ & $0.19 \pm 0.01$ & $445.21 \pm 5.17$ \\
 &  &  & 50 & $0.526000$ & $349.02 \pm 0.12$ & $7.97 \pm 0.50$ & $8.38 \pm 0.53$ & $0.58$ & $4.63 \pm 0.29$ & $428.70 \pm 1.98$ \\
 &  &  & 51 & $0.527000$ & $349.04 \pm 0.10$ & $5.17 \pm 2.54$ & $4.35 \pm 2.50$ & $0.85$ & $4.41 \pm 2.17$ & $432.62 \pm 0.87$ \\
 &  &  & 52 & $0.525000$ & $349.06 \pm 0.14$ & $3.85 \pm 2.25$ & $6.99 \pm 2.98$ & $0.26$ & $1.01 \pm 0.59$ & $369.12 \pm 4.98$ \\
 &  &  & 53 & $0.530000$ & $348.96 \pm 0.13$ & $2.92 \pm 0.26$ & $5.03 \pm 1.39$ & $0.28$ & $0.82 \pm 0.07$ & $334.18 \pm 2.80$ \\
 &  &  & 54 & $0.530000$ & $349.03 \pm 0.09$ & $6.12 \pm 1.09$ & $0.34 \pm 0.01$ & $0.25$ & $1.54 \pm 0.27$ & $392.01 \pm 4.80$ \\
 &  &  & 55 & $0.529000$ & $350.50 \pm 0.11$ & $9.50 \pm 2.67$ & $0.22 \pm 0.01$ & $0.23$ & $2.18 \pm 0.61$ & $336.27 \pm 2.24$ \\
\hline
12 Dec 2020 & 3.0 & $5$ & 56 & $0.538000$ & $350.40 \pm 0.13$ & $2.22 \pm 0.19$ & $9.65 \pm 6.06$ & $0.95$ & $2.11 \pm 0.18$ & $435.03 \pm 3.32$ \\
 &  &  & 57 & $0.540000$ & $347.81 \pm 0.14$ & $16.19 \pm 2.89$ & $1.55 \pm 0.29$ & $0.64$ & $10.30 \pm 1.84$ & $440.24 \pm 3.37$ \\
 &  &  & 58 & $0.545000$ & $345.50 \pm 0.17$ & $16.41 \pm 2.52$ & $1.58 \pm 0.28$ & $0.80$ & $13.07 \pm 2.01$ & $435.87 \pm 4.07$ \\
 &  &  & 59 & $0.543000$ & $341.24 \pm 0.12$ & $5.35 \pm 4.44$ & $10.66 \pm 7.66$ & $1.07$ & $5.73 \pm 4.76$ & $409.61 \pm 4.51$ \\
 &  &  & 60 & $0.547000$ & $347.11 \pm 0.14$ & $6.83 \pm 0.05$ & $11.53 \pm 5.61$ & $0.65$ & $4.42 \pm 0.03$ & $419.51 \pm 2.12$ \\
\enddata
\end{deluxetable*}
\begin{deluxetable*}{lcccccccccc}
\tablenum{2}
\label{tab:burst_parameters_2}
\tablecaption{Continued from Table 1.}
\tablewidth{\textwidth}
\tabletypesize{\scriptsize}
\tablehead{
\colhead{Epoch} & 
\colhead{Duration} & 
\colhead{No. of} & 
\colhead{Burst} & 
\colhead{Phase} & 
\colhead{$DM_{phase} \pm \delta DM_{phase}$} & 
\colhead{$w_{int} \pm \delta w_{int}$} & 
\colhead{$\tau \pm \delta \tau$} & 
\colhead{$S_\nu$} & 
\colhead{$F_{int} \pm \delta F_{int}$} & 
\colhead{$\nu_{centre} \pm \delta \nu_{centre}$} \\
\colhead{} & 
\colhead{(hour)} & 
\colhead{bursts} & 
\colhead{ID} & 
\colhead{} & 
\colhead{(pc cm$^{-3}$)} & 
\colhead{(ms)} & 
\colhead{(ms)} & 
\colhead{(Jy)} & 
\colhead{(Jy ms)} & 
\colhead{(MHz)} }
\startdata
26 Nov 2020 & $2.5$ & $14$ & 61 & $0.580000$ & $347.65 \pm 0.22$ & $8.80 \pm 2.04$ & $0.62 \pm 0.01$ & $0.40$ & $3.53 \pm 0.82$ & $348.30 \pm 10.32$ \\
 &  &  & 62 & $0.580000$ & $342.13 \pm 0.22$ & $3.89 \pm 1.93$ & $0.21 \pm 0.01$ & $0.67$ & $2.61 \pm 1.30$ & $315.10 \pm 0.96$ \\
 &  &  & 63 & $0.581000$ & $354.31 \pm 0.02$ & $19.99 \pm 1.25$ & $8.83 \pm 1.56$ & $0.76$ & $15.23 \pm 0.95$ & $415.08 \pm 3.79$ \\
 &  &  & 64 & $0.580000$ & $345.41 \pm 0.17$ & $4.50 \pm 0.79$ & $0.99 \pm 0.01$ & $0.73$ & $3.29 \pm 0.58$ & $382.44 \pm 5.66$ \\
 &  &  & 65 & $0.574000$ & $349.00 \pm 0.13$ & $24.89 \pm 8.01$ & $0.30 \pm 0.01$ & $0.54$ & $13.45 \pm 4.33$ & $376.35 \pm 3.22$ \\
 &  &  & 66 & $0.575000$ & $352.09 \pm 0.10$ & $4.79 \pm 1.99$ & $0.41 \pm 0.01$ & $0.30$ & $1.45 \pm 0.60$ & $454.90 \pm 1.19$ \\
 &  &  & 67 & $0.575000$ & $346.01 \pm 0.24$ & $7.29 \pm 0.50$ & $21.40 \pm 10.24$ & $0.63$ & $4.59 \pm 0.32$ & $365.05 \pm 0.46$ \\
 &  &  & 68 & $0.574000$ & $351.50 \pm 0.11$ & $19.05 \pm 4.47$ & $0.41 \pm 0.01$ & $0.47$ & $8.99 \pm 2.11$ & $372.86 \pm 1.36$ \\
 &  &  & 69 & $0.574000$ & $346.37 \pm 0.17$ & $11.73 \pm 1.64$ & $0.88 \pm 0.01$ & $0.49$ & $5.79 \pm 0.81$ & $385.17 \pm 4.64$ \\
 &  &  & 70 & $0.577000$ & $350.65 \pm 0.20$ & $29.53 \pm 4.20$ & $1.00 \pm 0.01$ & $0.43$ & $12.70 \pm 1.81$ & $435.24 \pm 3.17$ \\
 &  &  & 71 & $0.577000$ & $348.17 \pm 0.17$ & $12.42 \pm 6.39$ & $0.18 \pm 0.01$ & $0.42$ & $5.28 \pm 2.72$ & $412.53 \pm 2.63$ \\
 &  &  & 72 & $0.578000$ & $348.94 \pm 0.16$ & $29.11 \pm 3.85$ & $0.75 \pm 0.01$ & $0.41$ & $12.04 \pm 1.59$ & $418.98 \pm 0.17$ \\
 &  &  & 73 & $0.577000$ & $349.57 \pm 0.11$ & $10.80 \pm 3.66$ & $1.09 \pm 0.66$ & $0.45$ & $4.88 \pm 1.66$ & $353.43 \pm 3.68$ \\
 &  &  & 74 & $0.577000$ & $348.09 \pm 0.14$ & $11.85 \pm 3.48$ & $0.53 \pm 0.01$ & $0.54$ & $6.36 \pm 1.87$ & $416.92 \pm 4.49$ \\
\enddata
\end{deluxetable*}
\begin{deluxetable*}{cccccccc}
\tablenum{3}
\label{tab:literature_P1}
\tablecaption{The observed ranges of parameters of the bursts from $\FRB$ using different telescopes are summarised here. The symbols $DM$, $w_{int}$, and $\tau$ represent the dispersion measure, intrinsic width, and scattering width respectively.}
\tablewidth{0pt}
\tabletypesize{\normalsize}
\tablehead{
\colhead{Telescope} & \colhead{Frequency} & \colhead{No. of} & \colhead{Phase} & \colhead{DM range} & 
\colhead{$w_{int}$ range} & \colhead{$\tau$ range} & \colhead{Reference} \\
\colhead{Name} & \colhead{Range (MHz)} & \colhead{Bursts} & \colhead{Coverage} & \colhead{(pc cm$^{-3}$)} & 
\colhead{(ms)} & \colhead{(ms)} & \colhead{}
}
\startdata
CHIME	&	$400-800$	&	$38$	&	$-$	&	$348.8-350.1$	&	$0.58-8.60$	&	$-$	& \citealt{amiri20} \\
Effelsberg	&	$4000-8000$	&   $8$	&	$0.25-0.30$	& $-$	&	$0.26-3.42$	&	$-$	&\citealt{bethapudi23} \\
GBT	&	$300-400$	&	$8$	&	$0.47-0.57$	&	$348.7-349.5$	&	$1.50-5.89$	&	$1.80-5.90$	&	\citet{chawla20} \\
CHIME	&	$400-800$	&	$44$	&	$0.40-0.65$	&	$348.7-350.1$	&	$-$	&	$-$	&	\citet{mckinven23} \\
Apertif	&	$1220-1520$	&	$9$	&	$-$	&	$348.9-349.4$	&	$-$	&	$36.20-48.20$	&	\citet{pastor2021} \\
LOFAR	&	$110-190$	&	$54$	&	$-$	&	$347.3-350.7$	&	$-$	&	$-$	&	\citet{pastor2021}	\\
EVN	&	$1636-1764$	&	$4$	&	$-$	&	$345.0-356.0$	&	$0.24-1.86$	&	$-$	&	\citet{marcote20} \\
uGMRT	&	$550-750$	&	$15$	&	$-$	&	$347.8-349.0$	&	$-$	&	$-$	&	\citet{marthi20} \\
EVN	&	$1636-1764$	&	$4$	&	$-$	&	$-$	&	$0.24-1.86$	&	$-$	&	\citet{nimmo21} \\
CHIME	&	$400-800$	&	$8$	&	$-$	&	$348.7-370.4$	&	$0.76-8.60$	&	$-$	&	\citet{pearlman20} \\
LOFAR	&	$110-188$	&	$18$	&	$0.53-0.79$	&	$-$	&	$42.00-119.00$	&	$46.69-54.14$	&	\citet{pleunis21} \\
GBT	&	$600-1000$	&	$7$	&	$0.37-0.64$	&	$348.6-350.2$	&	$2.80-10.30$	&	$-$	&	\citet{sand22} \\
uGMRT	&	$300-500$	&	$4$	&	$0.37-0.64$	&	$348.6-350.2$	&	$2.46-8.40$	&	$1.67-2.51$	&	\citet{sand22} \\
SRT	&	$300-400$	&	$14$	&	$0.47-0.62$	&	$-$	&	$9.00-23.00$	&	$-$	&	\citet{trudu23} \\
uGMRT	&	$300-500$	&	$7$	&	$0.62-0.63$	&	$-$	&	$6.00-20.80$	&	$-$	&	\citet{trudu23} \\
\hline
uGMRT	&	$250-500$	&	$74$	&	$0.40-0.60$	&	$341.24-356.71$	&	$0.68-29.53$	&	$0.03-14.47$	&	 Current Analysis \\
\enddata
\end{deluxetable*}
\begin{deluxetable*}{cccccccc}
\tablenum{4}
\label{tab:literature_P2}
\tablecaption{The observed range of parameters of the bursts from $\FRB$ using different telescopes is summarized here. The symbols $S_{\nu}$, $F$ and $\mathcal{R}$ represent the flux density, fluence, and burst rate respectively.}
\tablewidth{0pt}
\tabletypesize{\normalsize}
\tablehead{
\colhead{Telescope} & \colhead{Frequency} & \colhead{No. of} & \colhead{$S_{\nu}$ range} & \colhead{$F$ range} & 
\colhead{$\mathcal{R}$ (hr$^{-1}$)} & \colhead{Drift Rate} & \colhead{Ref} \\
\colhead{Name} & \colhead{Range (MHz)} & \colhead{Bursts} & \colhead{(Jy)} & \colhead{(Jy ms)} & \colhead{($>F$ Jy ms)} & \colhead{(MHz ms$^{-1}$)} & \colhead{}
}
\startdata
CHIME	&	$400-800$	&	$38$	&	$0.30-6.10$	&	$0.40-37.00$	&	$-$	&	$-$	&	
\citealt{amiri20} \\
Effelsberg	&	$4000-8000$	&   $8$	&	$0.18-1.84$	&	$0.08-1.49$	&	$22.8\,(>0.005)$	&	$-$	&	
\citealt{bethapudi23} \\
GBT	&	$300-400$	&	$8$	&	$0.98-3.58$	&	$5.20-48.90$	&	$1.8\,(>22.0)$	&	$-4.2$	&	
\citet{chawla20} \\
CHIME	&	$400-800$	&	$44$	&	$-$	&	$-$	&	$-$	&	$-$	&	\citet{mckinven23} \\
Apertif	&	$1220-1520$	&	$9$	&	$-$	&	$38.00-318.00$	&	$-$	&	$-$	&	
\citet{pastor2021} \\
LOFAR	&	$110-190$	&	$54$	&	$-$	&	$1.20-58.30$	&	$-$	&	$-6.6$	&	
\citet{pastor2021}	\\	
EVN	&	$1636-1764$	&	$4$	& $0.50-2.30$	&	$0.20-2.53$	&	$-$	&	$-$	&	
\citet{marcote20} \\
uGMRT	&	$550-750$	&	$15$	&	$0.17-23.65$	&	$0.09-47.80$	&	$-$	&	$-$	&	
\citet{marthi20} \\
EVN	&	$1636-1764$	&	$4$	&	$-$	&	$0.20-2.53$	&	$-$	&	$-$	&	
\citet{nimmo21} \\
CHIME	&	$400-800$	&	$8$	&	$0.40-6.10$	&	$0.40-16.30$	&	$-$	&	$-$	&	
\citet{pearlman20} \\
LOFAR	&	$110-188$	&	$18$	&	$1.65-10.57$	& $26.00-308.00$	&	$0.3\,(>26.0)$	&	$-$	&	\citet{pleunis21} \\
GBT	&	$600-1000$	&	$7$	&	$0.12-4.39$	&	$0.24-6.10$	&	$4.2\,(>0.2)$	&	$-8.0$	&	
\citet{sand22} \\
uGMRT	&	$300-500$	&	$4$	&	$0.43-6.92$	&	$1.49-7.20$	& $0.7\,(>0.5)$	&	$-$	&	
\citet{sand22} \\
SRT	&	$300-400$	&	$14$	&	$2.10-12.2$	&	$15.00-172.00$	&	$-$	&	$-9.7$	&	
\citet{trudu23} \\
uGMRT	&	$300-500$	&	$7$	&	$0.20-7.90$	&	$0.60-7.90$	&	$-$	&	$-5.9$	&	
\citet{trudu23} \\
\hline
uGMRT	&	$250-500$	&	$74$	&	$0.12-2.40$	&	$0.14-15.23$	&	$4.0\,(>0.14)$	&	$-$	&	 
Current Analysis \\
\enddata
\end{deluxetable*}

\bibliography{references}

\begin{thebibliography}{}
\expandafter\ifx\csname natexlab\endcsname\relax\def\natexlab#1{#1}\fi
\providecommand{\url}[1]{\href{#1}{#1}}
\providecommand{\dodoi}[1]{doi:~\href{http://doi.org/#1}{\nolinkurl{#1}}}
\providecommand{\doeprint}[1]{\href{http://ascl.net/#1}{\nolinkurl{http://ascl.net/#1}}}
\providecommand{\doarXiv}[1]{\href{https://arxiv.org/abs/#1}{\nolinkurl{https://arxiv.org/abs/#1}}}

\bibitem[{Andersen {et~al.}(2019)Andersen, Bandura, Bhardwaj, Boubel, Boyce, Boyle, Brar, Cassanelli, Chawla, Cubranic, {et~al.}}]{andersen2019}
Andersen, B., Bandura, K., Bhardwaj, M., {et~al.} 2019, The Astrophysical Journal Letters, 885, L24

\bibitem[{Bethapudi {et~al.}(2024)Bethapudi, Spitler, Li, Marthi, Bause, Main, \& Wharton}]{bethapudi24}
Bethapudi, S., Spitler, L., Li, D., {et~al.} 2024, arXiv preprint arXiv:2409.12584

\bibitem[{Bethapudi {et~al.}(2023{\natexlab{a}})Bethapudi, Spitler, Main, Li, \& Wharton}]{bethapudi23}
Bethapudi, S., Spitler, L., Main, R., Li, D., \& Wharton, R. 2023{\natexlab{a}}, Monthly Notices of the Royal Astronomical Society, 524, 3303

\bibitem[{Bethapudi {et~al.}(2023{\natexlab{b}})Bethapudi, Spitler, Main, Li, \& Wharton}]{bethapudi2023}
---. 2023{\natexlab{b}}, Monthly Notices of the Royal Astronomical Society, 524, 3303

\bibitem[{Bhat {et~al.}(2004)Bhat, Cordes, Camilo, Nice, \& Lorimer}]{bhat04}
Bhat, N. D.~R., Cordes, J.~M., Camilo, F., Nice, D.~J., \& Lorimer, D.~R. 2004, The Astrophysical Journal, 605, 759

\bibitem[{Braga {et~al.}(2024)Braga, Cruces, Cassanelli, Espinoza-Dupouy, Rodriguez, Spitler, Vera-Casanova, \& Limaye}]{braga2024}
Braga, C., Cruces, M., Cassanelli, T., {et~al.} 2024, arXiv preprint arXiv:2408.12567

\bibitem[{Caleb {et~al.}(2019)Caleb, Stappers, Rajwade, \& Flynn}]{caleb2019}
Caleb, M., Stappers, B., Rajwade, K., \& Flynn, C. 2019, Monthly Notices of the Royal Astronomical Society, 484, 5500

\bibitem[{Chawla {et~al.}(2020)Chawla, Andersen, Bhardwaj, Fonseca, Josephy, Kaspi, Michilli, Pleunis, Bandura, Bassa, {et~al.}}]{chawla20}
Chawla, P., Andersen, B., Bhardwaj, M., {et~al.} 2020, The Astrophysical Journal Letters, 896, L41

\bibitem[{CHIME/FRB(2020)}]{amiri20}
CHIME/FRB. 2020, Nature, 582, 351

\bibitem[{CHIME/FRB-Collaboration {et~al.}(2021)CHIME/FRB-Collaboration, Amiri, Andersen, Bandura, {et~al.}}]{chime2021}
CHIME/FRB-Collaboration, Amiri, M., Andersen, B.~C., Bandura, K., {et~al.} 2021, The Astrophysical Journal Supplement Series, 257, 59

\bibitem[{CHIME/FRB-Collaboration {et~al.}(2020)CHIME/FRB-Collaboration, Amiri, Andersen, Bandura, M., {et~al.}}]{chime2020a}
CHIME/FRB-Collaboration, Amiri, M., Andersen, B.~C., {et~al.} 2020, Nature, 582, 351

\bibitem[{Connor {et~al.}(2020)Connor, Miller, \& Gardenier}]{connor2020}
Connor, L., Miller, M., \& Gardenier, D. 2020, Monthly Notices of the Royal Astronomical Society, 497, 3076

\bibitem[{Cruces {et~al.}(2021)Cruces, Spitler, Scholz, Lynch, Seymour, Hessels, Gouiff{\'e}s, Hilmarsson, Kramer, \& Munjal}]{cruces2021}
Cruces, M., Spitler, L., Scholz, P., {et~al.} 2021, Monthly Notices of the Royal Astronomical Society, 500, 448

\bibitem[{Deneva {et~al.}(2016)Deneva, Stovall, McLaughlin, Bagchi, Bates, Freire, Martinez, Jenet, \& Garver-Daniels}]{deneva2016}
Deneva, J., Stovall, K., McLaughlin, M., {et~al.} 2016, The Astrophysical Journal, 821, 10

\bibitem[{Gajjar {et~al.}(2018)Gajjar, Siemion, {et~al.}}]{gajjar18}
Gajjar, V., Siemion, A., {et~al.} 2018, ApJ, 863, 2

\bibitem[{Gupta {et~al.}(2017)Gupta, Ajithkumar, Kale, Nayak, Sabhapathy, Sureshkumar, Swami, Chengalur, Ghosh, Ishwara-Chandra, {et~al.}}]{gupta17}
Gupta, Y., Ajithkumar, B., Kale, H.~S., {et~al.} 2017, Current Science, 707

\bibitem[{Houben {et~al.}(2019{\natexlab{a}})Houben, Spitler, ter Veen, Rachen, Falcke, \& Kramer}]{houben2019}
Houben, L., Spitler, L., ter Veen, S., {et~al.} 2019{\natexlab{a}}, Astronomy \& Astrophysics, 623, A42

\bibitem[{Houben {et~al.}(2019{\natexlab{b}})Houben, Spitler, Ter~Veen, Rachen, Falcke, \& Kramer}]{houben19}
Houben, L. J.~M., Spitler, L.~G., Ter~Veen, S., {et~al.} 2019{\natexlab{b}}, Astronomy \& Astrophysics, 623, A42

\bibitem[{Kaplan \& Meier(1958)}]{kaplan1958}
Kaplan, E.~L., \& Meier, P. 1958, Journal of the American statistical association, 53, 457

\bibitem[{Keane \& Petroff(2015)}]{keane15}
Keane, E.~F., \& Petroff, E. 2015, Monthly Notices of the Royal Astronomical Society, 447, 2852

\bibitem[{Kumar {et~al.}(2019)Kumar, Shannon, {et~al.}}]{kumar19a}
Kumar, P., Shannon, R., {et~al.} 2019, ApJL, 887, L30

\bibitem[{Lorimer(2011)}]{sigproc}
Lorimer, D. 2011, Astrophysics Source Code Library, ascl

\bibitem[{Lorimer \& Kramer(2005)}]{lorimer2005}
Lorimer, D.~R., \& Kramer, M. 2005, Handbook of pulsar astronomy, Vol.~4 (Cambridge university press)

\bibitem[{Lyutikov {et~al.}(2020)Lyutikov, Barkov, \& Giannios}]{lyutikov2020}
Lyutikov, M., Barkov, M.~V., \& Giannios, D. 2020, The Astrophysical Journal Letters, 893, L39

\bibitem[{Marcote {et~al.}(2020)Marcote, Nimmo, {et~al.}}]{marcote20}
Marcote, B., Nimmo, K., {et~al.} 2020, Nature, 577, 190

\bibitem[{Marcote {et~al.}(2017)Marcote, Paragi, Hessels, Keimpema, Van~Langevelde, Huang, Bassa, Bogdanov, Bower, Burke-Spolaor, {et~al.}}]{marcote2017}
Marcote, B., Paragi, Z., Hessels, J., {et~al.} 2017, The Astrophysical Journal Letters, 834, L8

\bibitem[{Marthi {et~al.}(2020)Marthi, Gautam, Li, Lin, Main, Naidu, Pen, \& Wharton}]{marthi20}
Marthi, V.~R., Gautam, T., Li, D., {et~al.} 2020, Monthly Notices of the Royal Astronomical Society: Letters, 499, L16

\bibitem[{Mckinven {et~al.}(2023)Mckinven, Gaensler, Michilli, Masui, Kaspi, Bhardwaj, Cassanelli, Chawla, Dong, Fonseca, {et~al.}}]{mckinven23}
Mckinven, R., Gaensler, B., Michilli, D., {et~al.} 2023, The Astrophysical Journal, 950, 12

\bibitem[{Nimmo {et~al.}(2021)Nimmo, Hessels, Keimpema, Archibald, Cordes, Karuppusamy, Kirsten, Li, Marcote, \& Paragi}]{nimmo21}
Nimmo, K., Hessels, J., Keimpema, A., {et~al.} 2021, Nature Astronomy, 5, 594

\bibitem[{Ocker {et~al.}(2022)Ocker, Cordes, Chatterjee, Niu, Li, McKee, Law, Tsai, Anna-Thomas, Yao, {et~al.}}]{ocker2022}
Ocker, S.~K., Cordes, J.~M., Chatterjee, S., {et~al.} 2022, The Astrophysical Journal, 931, 87

\bibitem[{Oppermann {et~al.}(2018)Oppermann, Yu, \& Pen}]{oppermann18}
Oppermann, N., Yu, H.-R., \& Pen, U.-L. 2018, Monthly Notices of the Royal Astronomical Society, 475, 5109

\bibitem[{Ott {et~al.}(1994)Ott, Witzel, Quirrenbach, Krichbaum, Standke, Schalinski, \& Hummel}]{ott1994}
Ott, M., Witzel, A., Quirrenbach, A., {et~al.} 1994, Astronomy and Astrophysics (ISSN 0004-6361), vol. 284, no. 1, p. 331-339, 284, 331

\bibitem[{Pastor-Marazuela {et~al.}(2021)Pastor-Marazuela, Connor, van Leeuwen, Maan, Ter~Veen, Bilous, Oostrum, Petroff, Straal, Vohl, {et~al.}}]{pastor2021}
Pastor-Marazuela, I., Connor, L., van Leeuwen, J., {et~al.} 2021, Nature, 596, 505

\bibitem[{Pearlman {et~al.}(2020)Pearlman, Majid, Prince, Nimmo, Hessels, Naudet, \& Kocz}]{pearlman20}
Pearlman, A.~B., Majid, W.~A., Prince, T.~A., {et~al.} 2020, The Astrophysical Journal Letters, 905, L27

\bibitem[{Petroff {et~al.}(2019)Petroff, Oostrum, Stappers, Bailes, Barr, Bates, Bhandari, Bhat, Burgay, Burke-Spolaor, {et~al.}}]{petroff19}
Petroff, E., Oostrum, L.~C., Stappers, B.~W., {et~al.} 2019, Monthly Notices of the Royal Astronomical Society, 482, 3109

\bibitem[{Pilia {et~al.}(2020)Pilia, Burgay, Possenti, Ridolfi, Gajjar, Corongiu, Perrodin, Bernardi, Naldi, Pupillo, {et~al.}}]{pilia2020}
Pilia, M., Burgay, M., Possenti, A., {et~al.} 2020, The Astrophysical Journal Letters, 896, L40

\bibitem[{Pleunis {et~al.}(2021{\natexlab{a}})Pleunis, Michilli, Bassa, Hessels, Naidu, Andersen, Chawla, Fonseca, Gopinath, Kaspi, {et~al.}}]{pleunis2021}
Pleunis, Z., Michilli, D., Bassa, C., {et~al.} 2021{\natexlab{a}}, The Astrophysical Journal Letters, 911, L3

\bibitem[{Pleunis {et~al.}(2021{\natexlab{b}})Pleunis, Michilli, Bassa, Hessels, Naidu, Andersen, Chawla, Fonseca, Gopinath, Kaspi, {et~al.}}]{pleunis21}
---. 2021{\natexlab{b}}, The Astrophysical Journal Letters, 911, L3

\bibitem[{Rajwade {et~al.}(2020)Rajwade, Mickaliger, {et~al.}}]{rajwade2020}
Rajwade, K., Mickaliger, M., {et~al.} 2020, MNRAS, 495, 3551

\bibitem[{Rajwade \& van Leeuwen(2024)}]{rajwade2024}
Rajwade, K.~M., \& van Leeuwen, J. 2024, Universe, 10, 158

\bibitem[{Ravi {et~al.}(2019)Ravi, Catha, D’Addario, Djorgovski, Hallinan, Hobbs, Kocz, Kulkarni, Shi, Vedantham, {et~al.}}]{ravi19}
Ravi, V., Catha, M., D’Addario, L., {et~al.} 2019, Nature, 572, 352

\bibitem[{Sand {et~al.}(2022)Sand, Faber, Gajjar, Michilli, Andersen, Joshi, Kudale, Pilia, Brzycki, Cassanelli, {et~al.}}]{sand22}
Sand, K.~R., Faber, J.~T., Gajjar, V., {et~al.} 2022, The Astrophysical Journal, 932, 98

\bibitem[{Sand {et~al.}(2023)Sand, Breitman, Michilli, Kaspi, Chawla, Fonseca, Mckinven, Nimmo, Pleunis, Shin, {et~al.}}]{sand2023}
Sand, K.~R., Breitman, D., Michilli, D., {et~al.} 2023, The Astrophysical Journal, 956, 23

\bibitem[{Seymour {et~al.}(2019)Seymour, Michilli, \& Pleunis}]{seymour2019}
Seymour, A., Michilli, D., \& Pleunis, Z. 2019, Astrophysics Source Code Library

\bibitem[{Trudu {et~al.}(2023)Trudu, Pilia, Nicastro, Guidorzi, Orlandini, Zampieri, Marthi, Ambrosino, Possenti, Burgay, {et~al.}}]{trudu23}
Trudu, M., Pilia, M., Nicastro, L., {et~al.} 2023, Astronomy \& Astrophysics, 676, A17

\bibitem[{Wang {et~al.}(2019)Wang, Zhang, Chen, \& Xu}]{wang2019}
Wang, W., Zhang, B., Chen, X., \& Xu, R. 2019, The Astrophysical Journal Letters, 876, L15

\bibitem[{Zhang(2023)}]{zhang2023}
Zhang, B. 2023, Reviews of Modern Physics, 95, 035005

\end{thebibliography}
\bibliographystyle{aasjournal}

\end{document}